\pdfoutput=1

\documentclass[11pt,twoside,a4paper,cmspaper,final,collab]{cms-tdr}
\def\svnVersion{338442}\def\cmsCernNoTag{CERN-PH-EP/2015-285}\def\cmsCernDate{\today}\def\cmsMessage{Published in the Journal of High Energy Physics as \href{http://dx.doi.org/10.1007/JHEP04(2016)005}{\doi{10.1007/JHEP04(2016)005.}}}

\begin{document}\cmsNoteHeader{HIG-14-028}

\hyphenation{had-ron-i-za-tion}
\hyphenation{cal-or-i-me-ter}
\hyphenation{de-vices}
\RCS$Revision: 330691 $
\RCS$HeadURL: svn+ssh://svn.cern.ch/reps/tdr2/papers/HIG-14-028/trunk/HIG-14-028.tex $
\RCS$Id: HIG-14-028.tex 330691 2016-03-04 18:47:55Z alverson $
\newcommand{\KD}{\ensuremath{\mathcal{D}^\text{kin}_\text{bkg}}\xspace}
\newcommand{\superKD}{\ensuremath{\mathcal{D}_\text{bkg}}\xspace}
\newcommand{\spinKD}{\ensuremath{\mathcal{D}_{J^P}}\xspace}
\newcommand{\VDj}{\ensuremath{\mathcal{D}_\text{jet}}\xspace}
\newcommand{\VDu}{\ensuremath{\mathcal{D}_{\pt}}\xspace}
\newcommand{\MassD}{\ensuremath{\mathcal{D}_\text{mass}}\xspace}
\newcommand{\LikMuOneD}{\ensuremath{\mathcal{L}_\mathrm{1D}(m_{4\ell})}\xspace}
\newcommand{\LikMuTwoD}{\ensuremath{\mathcal{L}_\mathrm{3D}(m_{4\ell},\KD)}\xspace}
\newcommand{\LikMuThreeD}{\ensuremath{\mathcal{L}_\mathrm{3D}(m_{4\ell},\KD,\VDj/\VDu)}\xspace}
\newcommand{\LikMu}{\LikMuThreeD}
\newcommand{\LikMass}{\ensuremath{\mathcal{L}_\mathrm{3D}(m_{4\ell},\MassD,\KD)}\xspace}
\newcommand{\LikSpin}{\ensuremath{\mathcal{L}_\mathrm{3D}(\superKD,\VDj/\VDu,\spinKD)}\xspace}
\newcommand{\strength}{\ensuremath{\mu_\mathrm{S}}\xspace}
\newcommand{\usedLumiA}{\ensuremath{5.1\fbinv}\xspace}
\newcommand{\usedLumiB}{\ensuremath{19.7\fbinv}\xspace}
\newcommand{\Hllll}{\ensuremath{\PH \to 4\ell}\xspace}
\newcommand{\Zllll}{\ensuremath{\PZ \to 4\ell}\xspace}
\newcommand{\ggH}{\ensuremath{\Pg \Pg \to \PH}\xspace}
\newcommand{\VH}{\ensuremath{\mathrm{V} \PH}\xspace}
\newcommand{\ZH}{\ensuremath{\PZ \PH}\xspace}
\newcommand{\WH}{\ensuremath{\PW \PH}\xspace}
\newcommand{\ttH}{\ensuremath{\PQt \PAQt \PH}\xspace}

\cmsNoteHeader{HIG-14-028}
\title{Measurement of differential and integrated fiducial cross sections for Higgs boson production in the four-lepton decay channel in $\Pp \Pp$ collisions at $\sqrt{s} = 7$ and $8\TeV$}

\date{\today}

\abstract{
Integrated fiducial cross sections for the production of four leptons via the $ \PH \to 4\ell$ decays ($\ell = \Pe$, $\mu$) are measured in pp collisions at $\sqrt{s} = 7$ and $8\TeV$. Measurements are performed with data corresponding to integrated luminosities of 5.1\fbinv at 7\TeV, and 19.7\fbinv at 8\TeV, collected with the CMS experiment at the LHC. Differential cross sections are measured using the 8\TeV data, and are determined as functions of the transverse momentum and rapidity of the four-lepton system, accompanying jet multiplicity, transverse momentum of the leading jet, and difference in rapidity between the Higgs boson candidate and the leading jet. A measurement of the $ \PZ \to 4\ell $ cross section, and its ratio to the $ \PH \to 4\ell $ cross section is also performed. All cross sections are measured within a fiducial phase space defined by the requirements on lepton kinematics and event topology. The integrated $ \PH \to 4\ell $ fiducial cross section is measured to be $0.56^{+0.67}_{-0.44}\stat\;{}^{+0.21}_{-0.06}\syst\unit{fb}$ at 7\TeV, and $1.11^{+0.41}_{-0.35}\stat\;{}^{+0.14}_{-0.10}\syst\unit{fb}$ at 8\TeV. The measurements are found to be compatible with theoretical calculations based on the standard model.
}

\hypersetup{
pdfauthor={CMS Collaboration},
pdftitle={Measurement of differential and integrated fiducial cross sections for Higgs boson production in the four-lepton decay channel in pp collisions at sqrt(s) = 7 and 8 TeV},
pdfsubject={CMS},
pdfkeywords={CMS, physics, Higgs, fiducial, cross section}}

\maketitle

\section{Introduction} \label{sec:IntroductionPAS}
The observation of a new boson consistent with the standard model (SM) Higgs boson~\cite{Englert:1964et,Higgs:1964ia,Higgs:1964pj,Guralnik:1964eu,Higgs:1966ev,Kibble:1967sv} was reported by the ATLAS and CMS collaborations in 2012~\cite{Aad:2012tfa, Obs:2012gu}.
Subsequent measurements confirmed that the properties of the new boson, such as its couplings and decay width, are indeed consistent with expectations for the SM Higgs boson~\cite{Chatrchyan:2013lba, Aad:2013wqa, CMS:2014ega, Aad:2015gba, Aad:2015zhl} (and references given therein).

In this paper we present measurements of the integrated and differential cross sections for the production of four leptons via the $\PH \to 4\ell$ decays ($\ell = \Pe$, $\mu$) in pp collisions at centre-of-mass energies of 7 and 8\TeV. All cross sections are measured in a restricted part of the phase space (fiducial phase space) defined to match the experimental acceptance in terms of the lepton kinematics and topological event selection. The $\Hllll$ denotes the Higgs boson decay to the four-lepton final state via an intermediate pair of neutral electroweak bosons. A similar study of the Higgs boson production cross section using the $\Hllll$ decay channel has already been performed by the ATLAS Collaboration~\cite{Aad:2014tca}, while measurements in the $\PH \to 2\gamma$ decay channel have been reported by both the ATLAS and CMS collaborations~\cite{Aad:2014lwa, CMS:HIG-14-016}.

The integrated fiducial cross sections are measured using $\Pp \Pp$ collision data recorded with the CMS detector at the CERN LHC corresponding to integrated luminosities of \usedLumiA~at 7\TeV and \usedLumiB~at 8\TeV. The measurement of the ratio of cross sections at 7 and 8\TeV is also performed. The differential fiducial cross sections are measured using just the 8\TeV data, due to the limited statistics of the 7\TeV data set. The cross sections are corrected for effects related to detector efficiency and resolution. The fiducial phase space constitutes approximately 42\% of the total available phase space, and there is no attempt to extrapolate the measurements to the full phase space. This approach is chosen to reduce the systematic uncertainty associated with the underlying model of the Higgs boson properties and production mechanism. The remaining dependence of each measurement on the model assumptions is determined and quoted as a separate systematic effect. Due to the strong dependence of the cross section times branching fraction on the mass of the Higgs boson ($m_{\PH}$) in the region around 125\GeV, the measurements are performed assuming a mass of $m_{\PH} = 125.0\GeV$, as measured by the CMS experiment using the $\Hllll$ and $\PH \to 2\gamma$ channels~\cite{CMS:2014ega}. This approach also allows an easier comparison of measurements with the theoretical estimations.

The differential fiducial cross sections are measured as a function of several kinematic observables that are sensitive to the Higgs boson production mechanism: transverse momentum and rapidity of the four-lepton system, transverse momentum of the leading jet, separation in rapidity between the Higgs boson candidate and the leading jet, as well as the accompanying jet multiplicity. In addition, measurements of the $\Zllll$ fiducial cross section, and of its ratio to the corresponding $\Hllll$ fiducial cross section are also performed using the 8 TeV data. These measurements provide tests of the SM expectations, and important validations of our understanding of the detector response and methodology used for the $\Hllll$ cross section measurement. The results of the $\Hllll$ cross section measurements are compared to theoretical calculations in the SM Higgs sector that offer up to next-to-next-to-leading-order (NNLO) accuracy in perturbative QCD, and up to next-to-leading-order (NLO) accuracy in perturbative electro-weak corrections.

All measurements presented in this paper are based on the experimental techniques used in previous measurements of Higgs boson properties in this final state~\cite{Chatrchyan:2013mxa, Khachatryan:2014kca}. These techniques include: algorithms for the online event selection, algorithms for the reconstruction, identification and calibration of electrons, muons and jets, as well as the approaches to the event selection and background estimation.

This paper is organized as follows. The CMS detector and experimental techniques are briefly described in Section \ref{sec:DetectorObjects}. The data sets and simulated samples used in the analysis are described in Section \ref{sec:DatasetsSimulationPAS}. The event selection and background modelling are presented in Section~\ref{sec:EventsBackground}. The fiducial phase space used for the measurements is defined in Section~\ref{sec:Fiducial}, while the procedure for extracting the integrated and differential cross sections is presented in Section~\ref{sec:Methodology}. Section~\ref{sec:SystematicsPAS} discusses the systematic uncertainties in the measurements. Section \ref{sec:Results} presents the results of all measurements and their comparison with the SM-based calculations.

\section{The CMS detector and experimental methods}  \label{sec:DetectorObjects}

The central feature of the CMS apparatus is a superconducting solenoid of 6\unit{m} internal diameter, providing a magnetic field of 3.8\unit{T}. Within the solenoid volume are a silicon pixel and strip tracker, a lead tungstate crystal electromagnetic calorimeter, and a brass and scintillator hadron calorimeter, each composed of a barrel and two endcap sections. Forward calorimetry extends the pseudorapidity coverage provided by the barrel and endcap detectors to $|\eta| < 5$. Muons are measured in gas-ionization detectors embedded in the steel flux-return yoke outside the solenoid. A more detailed description of the CMS detector, together with a definition of the coordinate system used and the relevant kinematic variables, can be found in Ref.~\cite{Chatrchyan:2008zzk}.

The reconstruction of particles emerging from each collision event is obtained via a particle-flow event reconstruction technique. The technique uses an optimized combination of all information from the CMS sub-detectors to identify and reconstruct individual particles in the collision event~\cite{CMS-PAS-PFT-09-001, CMS-PAS-PFT-10-002}. The particles are classified into mutually exclusive categories: charged hadrons, neutral hadrons, photons, muons, and electrons. Jets are reconstructed from the individual particles using the anti-$\kt$ clustering algorithm with a distance parameter of 0.5~\cite{antikt}, as implemented in the \textsc{fastjet} package~\cite{Cacciari:fastjet2, Cacciari:2011ma}. Energy deposits from the multiple pp interactions (pileup) and from the underlying event are subtracted when computing the energy of jets and isolation of reconstructed objects using the \textsc{FastJet} technique~\cite{Cacciari:2007fd,Cacciari:2008gn,Cacciari:2011ma}.

Details on the experimental techniques for the reconstruction, identification, and isolation of electrons, muons and jets, as well as on the efficiencies of these techniques can be found in Refs.~\cite{CMS-PAS-PFT-10-001, CMS-PAS-PFT-10-002, CMS-PAS-PFT-10-003, CMS:2011aa, Chatrchyan:2012xi, Chatrchyan:2013sba, Khachatryan:2015hwa}. Details on the procedure used to calibrate the leptons and jets in this analysis can be found in Ref.~\cite{Chatrchyan:2013mxa}.

\section{Data and simulation samples} \label{sec:DatasetsSimulationPAS}

The data set analyzed was collected by the CMS experiment in 2011 and 2012, and corresponds to integrated luminosities of \usedLumiA of 7\TeV collision data and \usedLumiB of 8\TeV collision data, respectively.
The set of triggers used to collect the data set is the same as the one used in previous measurements of Higgs boson properties in four-lepton final states~\cite{Chatrchyan:2013mxa, Khachatryan:2014kca}.

Descriptions of the SM Higgs boson production in the gluon fusion ($\ggH$) process are obtained using the \textsc{HRes 2.3}~\cite{deFlorian:2012mx, Grazzini:2013mca}, \textsc{Powheg V2}~\cite{powheg, powhegvv}, and \textsc{Powheg MiNLO HJ}~\cite{Hamilton:2012np} generators. The \textsc{HRes} generator is a partonic level Monte Carlo (MC) generator that provides a description of the $\ggH$ process at NNLO accuracy in perturbative QCD and next-to-next-to-leading-logarithmic (NNLL) accuracy in the resummation of soft-gluon effects at small transverse momenta~\cite{deFlorian:2012mx, Grazzini:2013mca}. Since the resummation is inclusive over the QCD radiation recoiling against the Higgs boson, \textsc{HRes} is considered for the estimation of fiducial cross sections that are inclusive in the associated jet activity. The \textsc{HRes} estimations are obtained by choosing the central values for the renormalization and factorization scales to be $m_{\PH}=125.0\GeV$. The \textsc{ Powheg} generator is a partonic level matrix-element generator that implements NLO perturbative QCD calculations and additionally provides an interface with parton shower programs. It provides a description of the $\ggH$ production in association with zero jets at NLO accuracy. For the purpose of this analysis, it has been tuned using the \POWHEG damping factor $hdump$ of $104.16\GeV$, to closely match the Higgs boson $\pt$ spectrum in the full phase space, as estimated by the \textsc{HRes} generator. This factor minimises emission of the additional jets in the limit of large $\pt$, and enhances the contribution from the Sudakov form factor as $\pt$ approaches zero~\cite{powheg, powhegvv}. The \textsc{Powheg MiNLO HJ} generator is an extension of the \textsc{Powheg V2} generator based on the \textsc{MiNLO} prescription~\cite{Hamilton:2012np} for the improved next-to-leading-logarithmic accuracy applied to the $\ggH$ production in association with up to one additional jet. It provides a description of the $\ggH$ production in association with zero jets and one jet at NLO accuracy, and the $\ggH$ production in association with two jets only at the leading-order (LO) accuracy. All the generators used to describe the $\ggH$ process take into account the finite masses of the bottom and top quarks. The description of the SM Higgs boson production in the vector boson fusion (VBF) process is obtained at NLO accuracy using the \POWHEG generator. The processes of SM Higgs boson production associated with gauge bosons (\VH) or top quark-antiquark pair ($\ttH$) are described at LO accuracy using \textsc{Pythia 6.4}~\cite{Sjostrand:2006za}. The MC samples simulated with these generators are normalized to the inclusive SM Higgs boson production cross sections and branching fractions that correspond to the SM calculations at NNLO and NNLL accuracy, in accordance with the LHC Higgs Cross Section Working Group recommendations~\cite{Heinemeyer:2013tqa}.The \POWHEG samples of the $\ggH$ and VBF processes are used together with the \PYTHIA samples of the \VH and \ttH processes to model the SM signal acceptance in the fiducial phase space and to extract the results of the fiducial cross section measurements following the method described in Section~\ref{sec:Methodology}. These samples, together with the \textsc{HRes} and \textsc{Powheg MiNLO HJ} samples of the alternative description of the $\ggH$ process, are used to compare the measurement results to the SM-based theoretical calculations in Section~\ref{sec:Results}.

In order to estimate the dependence of the measurement procedure on the underlying assumption for the Higgs boson production mechanism, we have used the set of MC samples for individual production mechanisms described in the previous paragraph. In addition, in order to estimate the dependence of the measurement on different assumptions of the Higgs boson properties, we have also simulated a range of samples that describe the production and decay of exotic Higgs-like resonances to the four-lepton final state. These include spin-zero, spin-one, and spin-two resonances with anomalous interactions with a pair of neutral gauge bosons ($ \PZ \PZ$, $ \PZ \gamma^{*}$, $\gamma^{*}\gamma^{*}$) described by higher-order operators, as discussed in detail in Ref.~\cite{Khachatryan:2014kca}. All of these samples are generated using the \POWHEG generator for the description of NLO QCD effects in the production mechanism, and \textsc{JHUGen}~\cite{Gao:2010qx,Bolognesi:2012mm,Anderson:2013afp} to describe the decay of these exotic resonances to four leptons including all spin correlations.

The MC event samples that are used to estimate the contribution from the background process $\Pg\Pg \to \PZ \PZ$ are simulated using \textsc{MCFM 6.7}~\cite{Campbell:2011bn}, while the background process $\PQq\PAQq \to 4\ell$ is simulated at NLO accuracy with the \POWHEG generator including $s$-, $t$-, and $u$-channel diagrams. For the purpose of the $\Zllll$ cross section measurements, we have also separately modelled contributions from the $t$- and $u$-channels of the $\PQq\PAQq~(\to \PZ \PZ^{*})\to 4\ell$ process at NLO accuracy with \POWHEG.

All the event generators described above take into account the initial- and final-state QED radiation (FSR) effects which can lead to the presence of additional hard photons in an event. Furthermore, the \POWHEG and \textsc{JHUGen} event generators take into account interference between all contributing diagrams in the $\Hllll$ process, including those related to the permutations of identical leptons in the $4\Pe$ and $4\mu$ final states. In the case of the LO, NLO, and NNLO generators, the sets of parton distribution functions (PDF) \textsc{CTEQ6L}~\cite{cteq66}, \textsc{CT10}~\cite{CT10}, and \textsc{MSTW2008}~\cite{MSTW08} are used, respectively.

All generated events are interfaced with \textsc{Pythia 6.4.26} Tune $Z2^{*}$ to simulate the effects of the parton shower, multi-parton interactions, and hadronization. The \textsc{Pythia 6.4.26} $Z2^{*}$ tune is derived from the $Z1$ tune~\cite{Field:2010bc}, which uses the \textsc{CTEQ5L} parton distribution set, whereas $Z2^{*}$ adopts \textsc{CTEQ6L}~\cite{Pumplin:2002vw}. The \textsc{HRes} generator does not provide an interface with programs that can simulate the effects of hadronization and multi-parton interactions. In order to account for these effects in the \textsc{HRes} estimations, the \textsc{HRes} generator is used to first reweight the \textsc{Powheg+JHUGen} events simulated without multi-parton interaction and hadronization effects in a phase space that is slightly larger than the fiducial phase space. After that, the multi-parton interaction and hadronization effects are simulated using \PYTHIA and the reweighted \textsc{Powheg+JHUGen} events. The reweighting is performed separately for each observable of interest for the differential, as well as for the integrated cross section measurements. This procedure effectively adds the non-perturbative effects to the \textsc{HRes} partonic level estimations.

The generated events are processed through a detailed simulation of the CMS detector based on \GEANTfour~\cite{Agostinelli:2002hh, GEANT} and are reconstructed with the same algorithms that are used for data analysis. The pileup interactions are included in simulations to match the distribution of the number of interactions per LHC bunch crossing observed in data.  The average number of pileup interactions is measured to be approximately 9 and 21 in the 7 and 8\TeV data sets, respectively.

The selection efficiency in all the simulated samples is rescaled to correct for residual differences in lepton selection efficiencies in data and simulation. This correction is based on the total lepton selection efficiencies measured in inclusive samples of $\PZ$ boson events in simulation and data using a ``tag-and-probe'' method~\cite{CMS:2011aa}, separately for 7 and 8\TeV collisions. More details can be found in Ref.~\cite{Chatrchyan:2013mxa}.

\section{Event selection and background modelling} \label{sec:EventsBackground}

The measurements presented in this paper are based on the event selection used in the previous measurements of Higgs boson properties in this final state~\cite{Chatrchyan:2013mxa, Khachatryan:2014kca}. Events are selected online requiring the presence of a pair of electrons or muons, or a triplet of electrons. Triggers requiring an electron and a muon are also used. The minimum \pt of the leading and subleading lepton are 17 and 8\GeV, respectively, for the double-lepton triggers, while they are 15, 8 and 5\GeV for the triple-electron trigger. Events with at least four well identified and isolated electrons or muons are then selected offline, if they are compatible with being produced at the primary vertex. The primary vertex is selected to be the one with the highest sum of $\pt^2$ of associated tracks. Among all same-flavour and opposite-sign (SFOS) lepton pairs in the event, the one with an invariant mass closest to the nominal $\PZ$ boson mass is denoted $\PZ_{1}$  and retained if its mass, $m(\PZ_1)$, satisfies $40 \le m(\PZ_1) \le 120\GeV$. The remaining leptons are considered and the presence of a second $\ell^{+}\ell^{-}$ pair, denoted $\PZ_{2}$, is required with condition $12 \le m(\PZ_2) \le 120\GeV$. If more than one $\PZ_{2}$ candidate satisfies all criteria, the pair of leptons with the largest sum of the transverse momenta magnitudes, $\Sigma |\pt|$, is chosen. Among the four selected leptons $\ell_{i}$ ($i = 1..4$) forming the $\PZ_{1}$ and $\PZ_{2}$ candidates, at least one lepton should have $\pt \ge 20\GeV$, another one $\pt \ge 10\GeV$, and any opposite-charge pair of leptons $\ell_{i}^{+}$ and $\ell_{j}^{-}$, irrespective of flavor, must satisfy $m({\ell_{i}^{+}\ell_{j}^{-}}) \ge 4\GeV$. The algorithm to recover the photons from the FSR uses the same procedure as described in Ref.~\cite{Chatrchyan:2013mxa}.

In the analysis, the presence of jets is only used to determine the differential cross section measurements as a function of jet-related observables. Jets are selected if they satisfy $\pt \ge 30\GeV$ and  $|\eta| \le 4.7$, and are required to be separated from the lepton candidates and from identified FSR photons by $\Delta R \equiv \sqrt{\smash[b]{(\Delta \eta)^{2}+(\Delta \phi)^{2}}} > 0.5$ (where $\phi$ is the azimuthal angle in radians)~\cite{Chatrchyan:2013mxa}.

After the event selection is applied, the dominant contribution to the irreducible background for the $\Hllll$ process originates from the $\PZ \PZ$ production via the $\PQq \PAQq$ annihilation, while the subdominant contribution arises from the $\PZ \PZ$ production via gluon fusion. In those processes, at least one of the intermediate $\PZ$ bosons is not on-shell. The reducible backgrounds mainly arise from the processes where parts of intrinsic jet activity are misidentified as an electron or a muon, such as: production of $\PZ$ boson in association with jets, production of a $\PZ \PW$ boson pair in association with jets, and the $\ttbar$ pair production. Hereafter, this background is denoted as $\PZ+\mathrm{X}$. The other background processes have negligible contribution.

In the case of the $\Hllll$ cross section measurements, the irreducible $\PQq\PAQq \to \Z\Z$ and $\Pg\Pg \to \Z\Z$ backgrounds are evaluated from simulation based on generators discussed in Section~\ref{sec:DatasetsSimulationPAS}, following Ref.~\cite{Chatrchyan:2013mxa}. In the case of the $\Pg\Pg \to \Z\Z$ background, the LO cross section of $\Pg\Pg \to \Z\Z$ is corrected via a $m_{4\ell}$ dependent k-factor, as recommended in the study of Ref.~\cite{Bonvini:2013}.

\begin{table}[!hb]
\centering
\topcaption{
The number of estimated background and signal events, as well as the number of
observed candidates, after final inclusive selection in the range
$105 < m_{4\ell} < 140\GeV$, used in the $\Hllll$ measurements. Signal and $\PZ \PZ$
background are estimated from simulations, while the $\PZ+\mathrm{X}$ background
is evaluated using control regions in data.
\label{tab:EventYieldsH4l_PAS}}
\begin{tabular}{l|ccc}
\hline
Channel & $4\Pe$ & $4\Pgm$ & $2\Pe2\Pgm$ \\
\hline
\multicolumn{4}{c}{}  \\[-0.35cm]
\multicolumn{4}{c}{\usedLumiA (7\TeV)} \\
\hline
$\PQq\PAQq \to \PZ\PZ$ &  0.8  $\pm$  0.1  &  1.8  $\pm$  0.1  &  2.2  $\pm$  0.3  \\
$\PZ+\mathrm{X}$                & 0.3 $\pm$ 0.1     & 0.2 $\pm$ 0.1     & 1.0 $\pm$ 0.3\\
$\Pg\Pg \to \PZ\PZ$  &  0.03  $\pm$  0.01  &  0.06  $\pm$  0.02  &  0.07  $\pm$  0.02  \\
\hline
Total background expected & 1.2 $\pm$ 0.1 & 2.1 $\pm$ 0.1 &  3.4 $\pm$ 0.4 \\
\hline
$\Hllll$ ($m_{\PH} = 125.0\GeV$) &  0.7  $\pm$  0.1  &  1.2  $\pm$  0.1  & 1.7  $\pm$  0.3  \\
\hline
Observed  & 1 & 3 & 6 \\
\hline
\multicolumn{4}{c}{}  \\[-0.35cm]
\multicolumn{4}{c}{\usedLumiB (8\TeV)} \\
\hline
$\PQq\PAQq \to \PZ\PZ$ &  3.0 $\pm$  0.4  &  7.6  $\pm$  0.5  &  9.0  $\pm$  0.7 \\
$\PZ + \mathrm{X}$ & 1.5 $\pm$ 0.3 & 1.2 $\pm$ 0.5 & 4.2 $\pm$ 1.1 \\
$\Pg\Pg \to \PZ\PZ$ &  0.2  $\pm$  0.1  &  0.4  $\pm$  0.1  &  0.5  $\pm$  0.1  \\
\hline
Total background expected & 4.8 $\pm$ 0.7 & 9.2 $\pm$ 0.7 & 13.7 $\pm$ 1.3\phantom{0} \\
\hline
$\Hllll$ ($m_{\PH} = 125.0\GeV$) &  2.9 $\pm$  0.4  &  5.6  $\pm$  0.7  &  7.3  $\pm$  0.9  \\
\hline
Observed  & 9 & 15 & 15 \\
\hline
\end{tabular}
\end{table}

The reducible background ($\PZ + \mathrm{X}$) is evaluated using the method based on lepton misidentification probabilities and control regions in data,  following the procedure described in Ref.~\cite{Chatrchyan:2013mxa}. In the case of the integrated $\Hllll$ cross section measurement, the shape of the $m_{4\ell}$ distribution for the reducible background is obtained by fitting the $m_{4\ell}$ with empirical analytical functional forms presented in Ref.~\cite{Chatrchyan:2013mxa}. In the case of the differential $\Hllll$ measurements, the shapes of the reducible background are obtained from the control regions in data in the form of template functions, separately for each bin of the considered observable. The template functions are prepared following a procedure described in the spin-parity studies presented in Refs.~\cite{Chatrchyan:2013mxa, Khachatryan:2014kca}.

The number of estimated signal and background events for the $\Hllll$ measurement, as well as the number of observed candidates after the final inclusive selection in data in the mass region $105 < m_{4\ell} < 140\GeV$ are given in Table~\ref{tab:EventYieldsH4l_PAS}, separately for 7 and 8\TeV.

In part of the $m_{4\ell}$ spectrum below 100\GeV, the dominant contribution arises from the resonant $\Zllll$ production ($s$-channel of the $\PQq \PAQq \to 4\ell$ process via the $\PZ$ boson exchange). The sub-dominant contributions arise from the corresponding $t$- and $u$-channels of the $\PQq\PAQq \to 4\ell$ process, from the reducible background processes ($\PZ+\mathrm{X}$), as well as from the $\Pg\Pg \to \PZ\PZ$ background. In the case of the $\Zllll$ measurements, contributions from $s$-, $t$-, and $u$-diagrams of the $\PQq\PAQq \to 4\ell$ process (and their interference), and contribution of the $\Pg\Pg \to \PZ\PZ$ process are estimated from simulation. The $\PZ+\mathrm{X}$ background is evaluated using control regions in data following an identical procedure as the one described above. The expected number of events arising from the $s$-channel of the $\PQq\PAQq \to 4\ell$ process is $57.4 \pm 0.3$, from all other SM processes is $3.6 \pm 0.5$, and 72 candidate events are observed after the final inclusive selection in 8\TeV data in the mass region $50 < m_{4\ell} < 105\GeV$.

\begin{figure}[!hb]
\centering
      \includegraphics[width=0.48\textwidth]{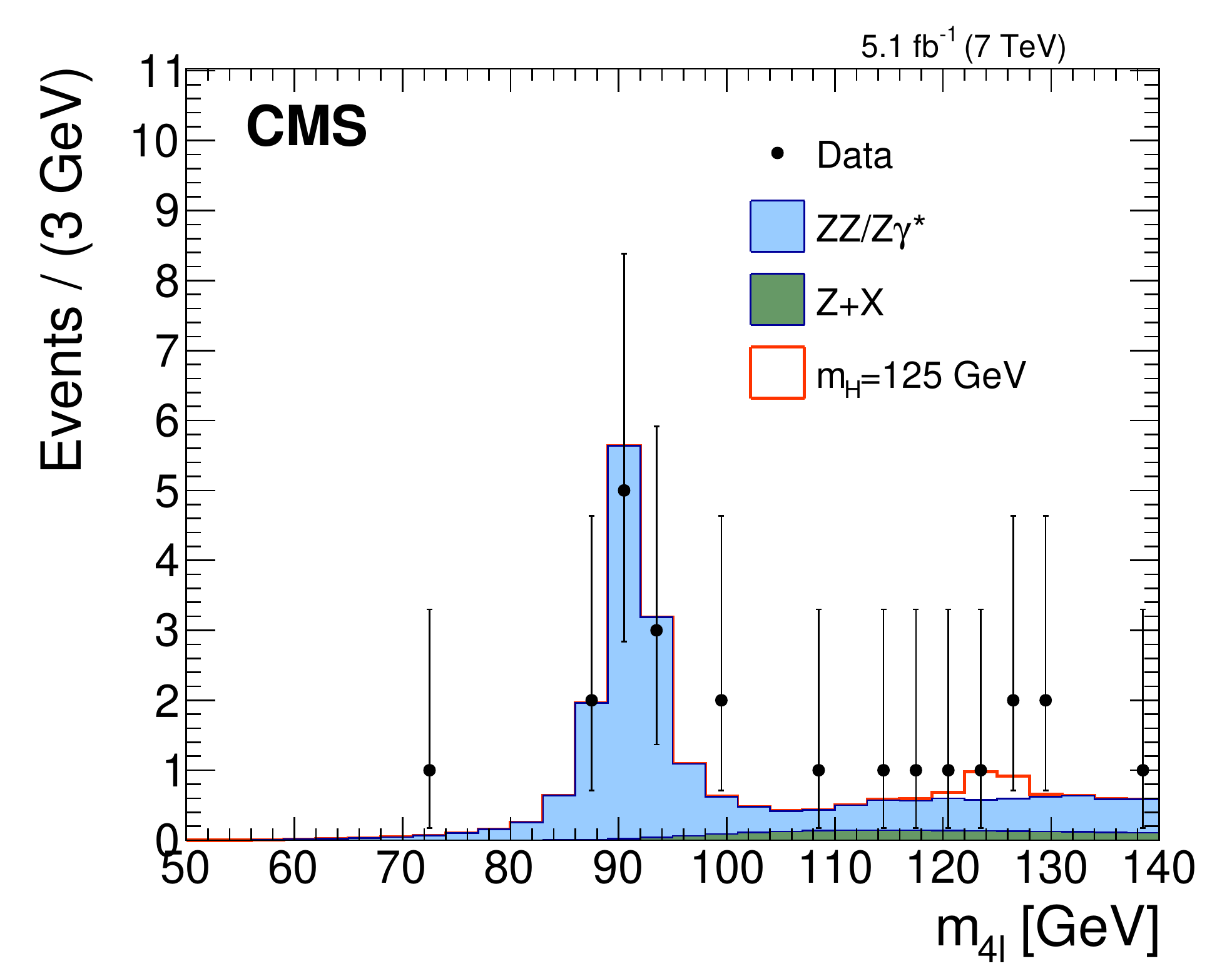}
      \includegraphics[width=0.48\textwidth]{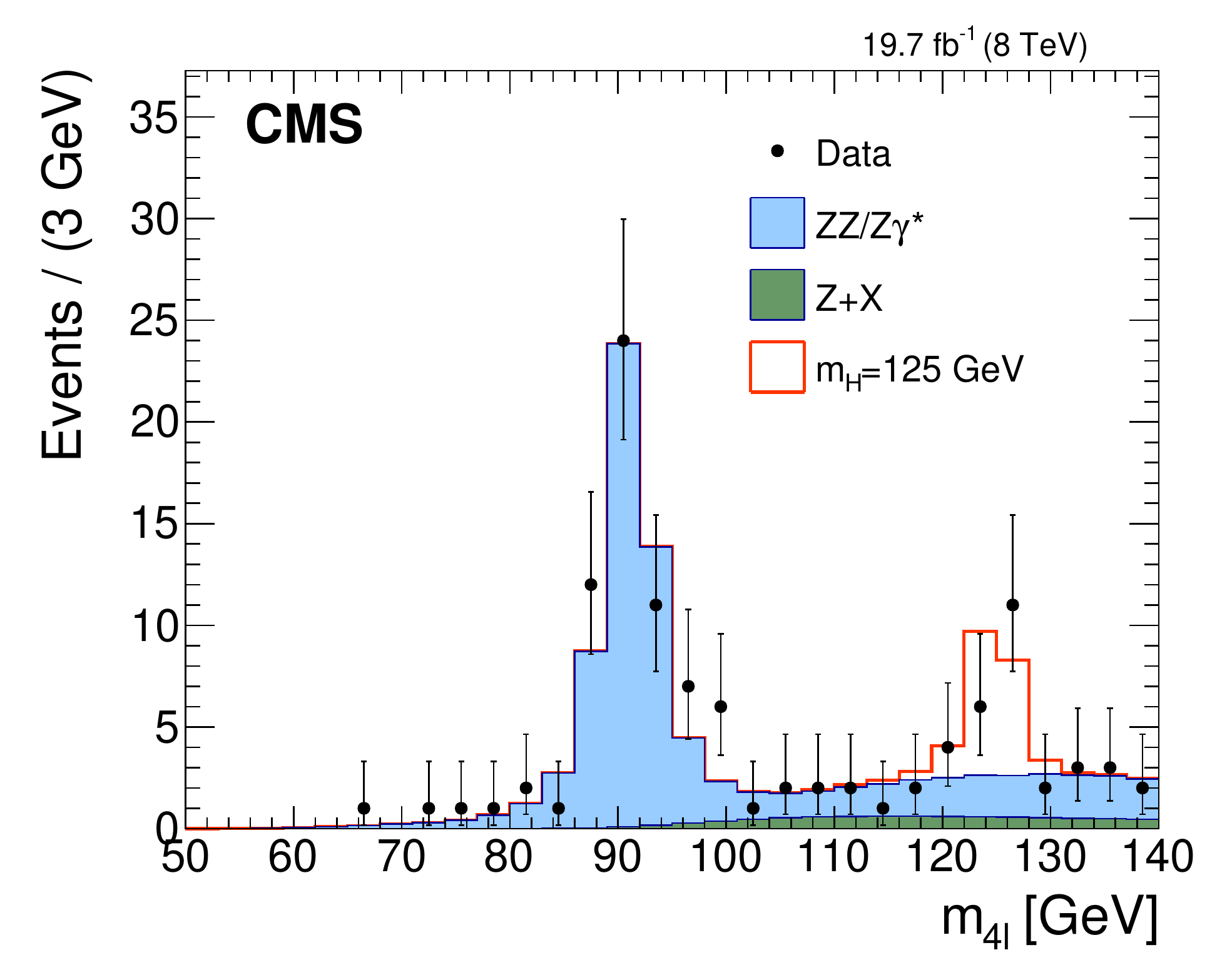}
    \caption{Distributions of the $m_{4\ell}$ observable in 7\TeV (left) and 8\TeV (right) data, as well as expectations for the SM Higgs boson ($m_{\PH}=125.0\GeV$) and other contributing SM processes, including resonant $\Zllll$ decays.}
  \label{fig:mass4l_PAS}
\end{figure}

The reconstructed four-lepton invariant mass distributions in the region of interest for the $\Hllll$ and $\Zllll$ measurements ($50 < m_{4\ell} < 140\GeV$) are shown in Fig.~\ref{fig:mass4l_PAS} for the 7 and 8\TeV data sets, and compared to the SM expectations.

\section{Fiducial phase space definition} \label{sec:Fiducial}

The acceptance and selection efficiency for the $\Hllll$ decays can vary significantly between different Higgs boson production mechanisms and different exotic models of Higgs boson properties. In processes with large jet activity (such as the $\ttH$ production), or with low invariant mass of the second lepton pair (such as $\PH \to  \PZ\gamma^{*}(\gamma^{*}\gamma^{*})  \to 4 \ell$ processes), or with the $\Hllll$ kinematics different from the SM estimation (such as exotic Higgs-like spin-one models), the inclusive acceptance of signal events can differ by up to 70\% from the inclusive acceptance estimated for SM $\Hllll$ decays.

In order to minimise the dependence of the measurement on the specific model assumed for Higgs boson production and properties, the fiducial phase space for the $\Hllll$ cross section measurements is defined to match as closely as possible the experimental acceptance defined by the reconstruction-level selection. This includes the definition of selection observables and selection requirements, as well as the definition of the algorithm for the topological event selection.

The fiducial phase space is defined using the leptons produced in the hard scattering, before any FSR occurs. This choice is motivated by the fact that the recovery of the FSR photons is explicitly performed at the reconstruction level. In the case of differential measurements as a function of jet-related observables, jets are reconstructed from the individual stable particles, excluding neutrinos and muons, using the anti-$k_t$ clustering algorithm with a distance parameter of 0.5. Jets are considered if they satisfy $\pt \ge 30\GeV$ and  $|\eta| \le 4.7$.

The fiducial phase space requires at least four leptons (electrons, muons), with at least one lepton having $\pt > 20\GeV$, another lepton having $\pt > 10\GeV$, and the remaining electrons and muons having $\pt > 7\GeV$ and $\pt > 5\GeV$ respectively.  All electrons and muons must have pseudorapidity $|\eta|<2.5$ and $|\eta|<2.4$, respectively. In addition, each lepton must satisfy an isolation requirement computed using the $\pt$ sum of all stable particles within $\Delta R<0.4$ distance from that lepton. The $\pt$ sum excludes any neutrinos, as well as any photon or stable lepton that is a daughter of the lepton for which the isolation sum is being computed. The ratio of this sum and the $\pt$ of the considered lepton must be less than $0.4$, in line with the requirement on the lepton isolation at the reconstruction level~\cite{Chatrchyan:2013mxa}. The inclusion of isolation is an important step in the fiducial phase space definition as it reduces significantly the differences in signal selection efficiency between different signal models.  It has been verified in simulation that the signal selection efficiency differs by up to 45\% between different models if the lepton isolation requirement is not included. This is especially pronounced in case of large associated jet activity as in the case of $\ttH$ production mode.
Exclusion of neutrinos and FSR photons from the computation of the isolation sum brings the definition of the fiducial phase space closer to the reconstruction level, and improves the model independence of the signal selection efficiency by an additional few percent.

Furthermore, an algorithm for a topological selection closely matching the one at the reconstruction level is applied as part of the fiducial phase space definition. At least two SFOS lepton pairs are required, and all SFOS lepton pairs are used to form $\PZ$ boson candidates. The SFOS pair with invariant mass closest to the nominal $\PZ$ boson mass ($91.188\GeV$) is taken as the first $\PZ$ boson candidate (denoted as $\PZ_{1}$). The mass of the $\PZ_{1}$ candidate must satisfy $40 < m(\PZ_{1}) < 120\GeV$. The remaining set of SFOS pairs are used to form the second $\PZ$ boson candidate (denoted as $\PZ_{2}$). In events with more than one $\PZ_{2}$ candidate, the SFOS pair with the largest sum of the transverse momenta magnitudes, $\Sigma |\pt|$, is chosen. The mass of the $\PZ_{2}$ candidate must satisfy $12 < m(\PZ_{2}) < 120\GeV$. Among the four selected leptons, any pair of leptons $\ell_{i}$ and $\ell_{j}$ must satisfy $\Delta R(\ell_{i}\ell_{j})>0.02$. Similarly, of the four selected leptons, the invariant mass of any opposite-sign lepton pair must satisfy $m(\ell_{i}^{+}\ell_{j}^{-})>4\GeV$.
Finally, the invariant mass of the Higgs boson candidate must satisfy $105 < m_{4\ell} < 140\GeV$.
The requirement on the $m_{4\ell}$ is important as the off-shell production cross section in the dominant gluon fusion production mode is sizeable and can amount up to a few percent of the total cross section~\cite{Kauer:2012hd}.
All the requirements and selections used in the definition of the fiducial phase space are summarised in Table~\ref{tab:FidDef}.

\begin{table}[!h!tb]
\centering
\topcaption{
Summary of requirements and selections used in the definition of the fiducial phase space for the $\Hllll$ cross section measurements. For measurements of the $\Zllll$ cross section and the ratio of the $\Hllll$ and $\Zllll$ cross sections, the requirement on the invariant mass of the selected four leptons is modified accordingly. More details, including the exact definition of the stable particles and lepton isolation, as well as $\PZ_1$ and $\PZ_2$ candidates, can be found in the text.
\label{tab:FidDef}
}
\begin{tabular}{lr}
\hline
\multicolumn{2}{c}{Requirements for the $\Hllll$ fiducial phase space} \\
\hline
\multicolumn{2}{c}{}  \\[-0.3cm]
\multicolumn{2}{c}{Lepton kinematics and isolation} \\
\hline
Leading lepton $\pt$ & $\pt  > 20\GeV$ \\
Sub-leading lepton $\pt$ & $\pt  > 10\GeV$ \\
Additional electrons (muons) $\pt$ & $\pt  > 7~(5)\GeV$ \\
Pseudorapidity of electrons (muons) & $|\eta| < 2.5~(2.4)$ \\
Sum of scalar $\pt$ of all stable particles within $\Delta R < 0.4$ from lepton & $< 0.4 \pt$ \\
\hline
\multicolumn{2}{c}{}  \\[-0.3cm]
\multicolumn{2}{c}{Event topology} \\
\hline
\multicolumn{2}{l}{Existence of at least two SFOS lepton pairs, where leptons satisfy criteria above} \\
Inv. mass of the $\PZ_1$ candidate & $40 < m(\PZ_{1})< 120\GeV$ \\
Inv. mass of the $\PZ_2$ candidate & $12 < m(\PZ_{2})< 120\GeV$ \\
Distance between selected four leptons & $\Delta R(\ell_{i}\ell_{j})>0.02$  \\
Inv. mass of any opposite-sign lepton pair & $m(\ell^{+}_{i}\ell^{-}_{j})>4\GeV$ \\
Inv. mass of the selected four leptons & $105 < m_{4\ell} < 140\GeV$  \\
\hline
\end{tabular}
\end{table}

It has been verified in simulation that the reconstruction efficiency for events originating from the fiducial phase space defined in this way only weakly depends on the Higgs boson properties and production mechanism. The systematic effect associated with the remaining model dependence is extracted and quoted separately, considering a wide range of alternative Higgs boson models, as described in Section~\ref{sec:SystematicsPAS}. The fraction of signal events within the fiducial phase space $\mathcal{A}_\text{fid}$, and the reconstruction efficiency $\epsilon$ for signal events within the fiducial phase space for individual SM production modes and exotic signal models are listed in Table~\ref{tab:Accept_Eff_fOut}.

\begin{table}[!h!tb]
\renewcommand{\arraystretch}{1.1}
\centering
\topcaption{
The fraction of signal events within the fiducial phase space (acceptance $\mathcal{A}_\text{fid}$), reconstruction efficiency ($\epsilon$) for signal events from within the fiducial phase space, and ratio of reconstructed events which are from outside the fiducial phase space to reconstructed events which are from within the fiducial phase space ($f_\text{nonfid}$). Values are given for characteristic signal models assuming $m_{\PH} = 125.0\GeV$, $\sqrt{s} = 8\TeV$, and the uncertainties include only the statistical uncertainties due to the finite number of events in MC simulation. In case of the first seven signal models, decays of the Higgs-like boson to four leptons proceed according to SM via the $\PH \to \PZ \PZ^{*} \to 4\ell$ process. Definition of signal excludes events where at least one reconstructed lepton originates from associated vector bosons or jets. The factor $(1+f_\text{nonfid})\epsilon$ is discussed in Section~\ref{sec:Methodology}.
\label{tab:Accept_Eff_fOut}
}
\resizebox{\textwidth}{!}{
\begin{tabular}{l|cccc}
\hline
Signal process & $\mathcal{A}_\text{fid}$ & $\epsilon$ & $f_\text{nonfid}$  & $(1+f_\text{nonfid})\epsilon$ \\
\hline
\multicolumn{2}{c}{}  \\[-0.35cm]
\multicolumn{5}{c}{Individual Higgs boson production modes} \\
\hline
$\ggH$ (\textsc{Powheg+JHUGen})   & 0.422 $\pm$ 0.001 & 0.647 $\pm$ 0.002 & 0.053 $\pm$ 0.001  & 0.681 $\pm$ 0.002 \\
VBF (\POWHEG)                      & 0.476 $\pm$ 0.003 & 0.652 $\pm$ 0.005 & 0.040 $\pm$ 0.002  & 0.678 $\pm$ 0.005 \\
WH (\PYTHIA)                          & 0.342 $\pm$ 0.002 & 0.627 $\pm$ 0.003 & 0.072 $\pm$ 0.002  & 0.672 $\pm$ 0.003 \\
ZH (\PYTHIA)                          & 0.348 $\pm$ 0.003 & 0.634 $\pm$ 0.004 & 0.072 $\pm$ 0.003  & 0.679 $\pm$ 0.005 \\
$\ttH$ (\PYTHIA)                      & 0.250 $\pm$ 0.003 & 0.601 $\pm$ 0.008 & 0.139 $\pm$ 0.008  & 0.685 $\pm$ 0.010 \\
\hline
\multicolumn{2}{c}{}  \\[-0.35cm]
\multicolumn{5}{c}{Some characteristic models of a Higgs-like boson with exotic decays and properties} \\
\hline
$\PQq\PAQq\to \PH(J^{CP}=1^{-})$ (\textsc{JHUGen})         & 0.238 $\pm$ 0.001 & 0.609 $\pm$ 0.002 & 0.054 $\pm$ 0.001  & 0.642 $\pm$ 0.002 \\
$\PQq\PAQq\to \PH(J^{CP}=1^{+})$ (\textsc{JHUGen})        & 0.283 $\pm$ 0.001 & 0.619 $\pm$ 0.002 & 0.051 $\pm$ 0.001  & 0.651 $\pm$ 0.002 \\
$\Pg\Pg \to \PH \to \PZ\gamma^{*}$ (\textsc{JHUGen})         & 0.156 $\pm$ 0.001 & 0.622 $\pm$ 0.002 & 0.073 $\pm$ 0.001  & 0.667 $\pm$ 0.002 \\
$\Pg\Pg \to \PH \to \gamma^{*}\gamma^{*}$ (\textsc{JHUGen}) & 0.188 $\pm$ 0.001 & 0.629 $\pm$ 0.002 & 0.066 $\pm$ 0.001  & 0.671 $\pm$ 0.002 \\
\hline
\end{tabular}
}
\end{table}

It should be noted that the cross section is measured for the process of resonant production of four leptons via the $\Hllll$ decays. This definition excludes events where at least one reconstructed lepton originates from associated vector bosons or jets, and not from the $\Hllll$ decays. Those events present a broad $m_{4\ell}$ distribution, whose exact shape depends on the production mode, and are treated as a combinatorial signal-induced background in the measurement procedure. This approach provides a simple measurement procedure with a substantially reduced signal model dependence. More details are discussed in Section~\ref{sec:Methodology}.

In the case of the independent measurement of the $\Zllll$ fiducial cross section, the fiducial phase space is defined in the analogous way, with the difference that the invariant mass of the $4\ell$ candidate for the $\PZ$ boson must satisfy $50 < m_{4\ell} < 105\GeV$. In the case of the measurement of the ratio of the $\Hllll$ and $\Zllll$ cross sections, the mass window of $50 < m_{4\ell} < 140\GeV$ is used.

\section{Measurement methodology} \label{sec:Methodology}

The aim is to determine the integrated and differential cross sections within the fiducial phase space, corrected for the effects of limited detection efficiencies, resolution, and known systematic biases. In order to achieve this goal, we estimate those effects using simulation and include them in the parameterization of the expected $m_{4\ell}$ spectra at the reconstruction level. We then perform a maximum likelihood fit of the signal and background parameterizations to the observed $4\ell$ mass distribution, $N_{\text{obs}}(m_{4\ell})$, and directly extract the fiducial cross sections of interest ($\sigma_{\mathrm{fid}}$) from the fit.  In this approach all systematic uncertainties are included in the form of nuisance parameters, which are effectively integrated out in the fit procedure. The results of measurements are obtained using an asymptotic approach~\cite{Cowan:2010js} with the test statistics based on the profile likelihood ratio~\cite{LHC-HCG}. The coverage of the quoted intervals obtained with this approach has been verified for a subset of results using the Feldman-Cousins method~\cite{Feldman:1997qc}. The maximum likelihood fit is performed simultaneously in all final states and in all bins of the observable considered in the measurement, assuming a Higgs boson mass of $m_{\PH} = 125.0\GeV$. The integrated cross section measurement is treated as a special case with a single bin. This implementation of the procedure for the unfolding of the detector effects from the observed distributions is different from the implementations commonly used in the experimental measurements, such as those discussed in Ref.~\cite{Prosper:2011zz}, where signal extraction and unfolding are performed in two separate steps. It is similar to the approach adopted in Ref.~\cite{CMS:HIG-14-016}.

The shape of the resonant signal contribution, $\mathcal{P}_{\text{res}}(m_{4\ell})$, is described by a double-sided Crystal Ball function as detailed in Ref.~\cite{Chatrchyan:2013mxa}, with a normalization proportional to the fiducial cross section $\sigma_{\mathrm{fid}}$. The shape of the combinatorial signal contribution, $\mathcal{P}_{\text{comb}}(m_{4\ell})$, from events where at least one of the four leptons does not originate from the $\Hllll$ decay, is empirically modelled by a Landau distribution whose shape parameters are constrained in the fit to be within a range determined from simulation. The remaining freedom in these parameters results in an additional systematic uncertainty on the measured cross sections. This contribution is treated as a background and hereafter we refer to this contribution as the ``combinatorial signal'' contribution. This component in the mass range $105 < m_{4\ell} < 140\GeV$ amounts to about 4\%, 18\%, and 22\% for $\WH$, $\ZH$, and $\ttH$ production modes, respectively.

An additional resonant signal contribution from events that do not originate from the fiducial phase space can arise due to detector effects that cause differences between the quantities used for the fiducial phase space definition, such as the lepton isolation, and the analogous quantities used for the event selection. This contribution is also treated as background, and hereafter we refer to this contribution as the ``nonfiducial signal'' contribution. It has been verified in simulation that the shape of these events is identical to the shape of the resonant fiducial signal and, in order to minimise the model dependence of the measurement, its normalization is fixed to be a fraction of the fiducial signal component. The value of this fraction, which we denote by $f_{\text{nonfid}}$, has been determined from simulation for each of the studied signal models, and it varies from ${\sim}5\%$ for the $\ggH$ production to ${\sim}14\%$ for the $\ttH$ production mode. The variation of this fraction between different signal models is included in the model dependence estimation. The value of $f_{\text{nonfid}}$ for different signal models is shown in Table~\ref{tab:Accept_Eff_fOut}.

In order to compare with the theoretical estimations, the measurement needs to be corrected for limited detector efficiency and resolution effects. The efficiency for an event passing the fiducial phase space selection to pass the reconstruction selection is measured using signal simulation samples and corrected for residual differences between data and simulation, as briefly described in Section~\ref{sec:DatasetsSimulationPAS} and detailed in Ref.~\cite{Chatrchyan:2013mxa}. It is determined from simulations that this efficiency for the $\ggH$ process is about 65\% inclusively, and that it can vary relative to the $\ggH$ process by up to ${\sim}7 \%$ in other signal models, as shown in Table~\ref{tab:Accept_Eff_fOut}. The largest deviations from the overall efficiency that correspond to the SM Higgs boson are found to be from $\ttH$ production, the H$\to \PZ\gamma^{*} \to 4 \ell$ process, and exotic Higgs-like spin-one models.

In the case of the differential cross section measurements, the finite efficiencies and resolution effects are encoded in a detector response matrix that describes how events migrate from a given observable bin at the fiducial level to a given bin at the reconstruction level. This matrix is diagonally dominant for the jet inclusive observables, but has sizeable off-diagonal elements for the observables involving jets. In the case of the jet multiplicity measurement the next-to-diagonal elements range from 3\% to 21\%, while in the case of other observables these elements are typically of the order of 1--2\%.

Following the models for signal and background contributions described above, the number of expected events in each final state $\mathrm{f}$ and in each bin $i$ of a considered observable is expressed as a function of $m_{4\ell}$ given by:
\begin{equation}
\label{eqn:m4l}
\begin{aligned}
N_{\text{obs}}^{\mathrm{f},i}(m_{4\ell}) =& N_{\mathrm{fid}}^{\mathrm{f},i}(m_{4\ell})+N_{\text{nonfid}}^{\mathrm{f},i}(m_{4\ell})+N_{\text{comb}}^{\mathrm{f},i}(m_{4\ell})+N_{\text{bkd}}^{\mathrm{f},i}(m_{4\ell}) \\
=&\sum_{j} \epsilon_{i,j}^{\mathrm{f}} \, \left(1+f_\text{nonfid}^{\mathrm{f},i} \right)\,\sigma_{\mathrm{fid}}^{\mathrm{f},j} \, \mathcal{L}\,\mathcal{P}_{\text{res}}(m_{4\ell}) \\
&+ N_{\text{comb}}^{\mathrm{f},i}\,\mathcal{P}_{\text{comb}}(m_{4\ell})+N_{\text{bkd}}^{\mathrm{f},i}\,\mathcal{P}_{\text{bkd}}(m_{4\ell}).
\end{aligned}
\end{equation}
The components $N_{\mathrm{fid}}^{\mathrm{f},i}(m_{4\ell})$, $N_{\text{nonfid}}^{\mathrm{f},i}(m_{4\ell})$, $N_{\text{comb}}^{\mathrm{f},i}(m_{4\ell})$, and $N_{\text{bkd}}^{\mathrm{f},i}(m_{4\ell})$ represent the resonant fiducial signal, resonant nonfiducial signal, combinatorial contribution from fiducial signal, and background contributions in bin $i$ as functions of $m_{4\ell}$, respectively. Similarly, the $\mathcal{P}_{\text{res}}(m_{4\ell})$, $\mathcal{P}_{\text{comb}}(m_{4\ell})$ and $\mathcal{P}_{\text{bkd}}(m_{4\ell})$ are the corresponding probability density functions for the resonant (fiducial and nonfiducial) signal, combinatorial signal, and background contributions. The $\epsilon_{i,j}^{\mathrm{f}}$ represents the detector response matrix that maps the number of expected events in a given observable bin $j$ at the fiducial level to the number of expected events in the bin $i$ at the reconstruction level. The $f_\text{nonfid}^{i}$ fraction describes the ratio of the nonfiducial and fiducial signal contribution in bin $i$ at the reconstruction level. The parameter $\sigma_{\mathrm{fid}}^{\mathrm{f},j}$ is the signal cross section for the final state $\mathrm{f}$ in bin $j$ of the fiducial phase space.

To extract the $4\ell$ fiducial cross-sections, $\sigma_{\mathrm{fid}}^{\mathrm{4\ell},j}$, in all bins $j$ of a considered observable, an unbinned likelihood fit is performed simultaneously for all bins $i$ at reconstruction level on the mass distributions of the three final states $4\Pe$, $4\mu$, and $2\Pe2\mu$, using Eq.~\eqref{eqn:m4l}. In each bin j of the fiducial phase space the fitted parameters are $\sigma_{\mathrm{fid}}^{\mathrm{4\ell},j}$, the sum of the three final state cross-sections, and two remaining degrees of freedom for the relative contributions of the three final states.

The inclusive values of the factor $(1+f_\text{nonfid})\epsilon$ from Eq.~\eqref{eqn:m4l} are shown in Table~\ref{tab:Accept_Eff_fOut} for different signal production modes and different exotic models. The relatively weak dependence of this factor on the exact signal model is a consequence of the particular definition of the fiducial phase space introduced in Section~\ref{sec:Fiducial}, and enables a measurement with a very small dependence on the signal model.

In the case of the simultaneous fit for the $\Hllll$ signal in 7 and 8\TeV data sets, and the measurement of the ratio of the $\Hllll$ cross sections at 7 and 8\TeV, the procedure described above is generalised to include two separate signals. The parameters extracted simultaneously from the measurement are the 8\TeV fiducial cross section, and ratio of 7\TeV and 8\TeV fiducial cross sections.

In the case of the $\Zllll$ cross section measurements, the definition of the fiducial phase space and statistical procedure are analogous to the ones used for the $\Hllll$ cross section measurements with the $\PZ$ boson mass fixed to the PDG value of $m_{\PZ} = 91.188\GeV$~\cite{PDG}.

Similarly, in the case of the simultaneous fit for the $\Hllll$ and $\Zllll$ signals, and the measurement of the ratio of the $\Hllll$ and $\Zllll$ cross sections, the procedure described above is generalised to include two separate signals. The parameters extracted simultaneously from this measurement are the $\Hllll$ fiducial cross section, and ratio of the $\Hllll$ and $\Zllll$ fiducial cross sections. Furthermore, this measurement is performed in two scenarios. In the first scenario, we fix the Higgs boson mass to $m_{\PH} = 125.0\GeV$ and the $\PZ$ boson mass to its PDG value. Results of measurements obtained in this scenario are reported in Section~\ref{sec:Results}. In the second scenario, we allow the masses of the two resonances to vary, and we fit for the mass of the Higgs boson $m_{\PH}$ and the mass difference between the two bosons $\Delta m = m_{\PH} - m_{\PZ}$. This scenario allows for an additional reduction of the systematic uncertainties related to the lepton momentum scale determination, and provides an additional validation of the measurement methodology.

\section{Systematic uncertainties}  \label{sec:SystematicsPAS}

Experimental systematic uncertainties in the parameterization of the signal and the irreducible background processes due to the trigger and combined lepton reconstruction, identification, and isolation efficiencies are evaluated from data and found to be in the range 4--10\%~\cite{Chatrchyan:2013mxa}. Theoretical uncertainties in the irreducible background rates are estimated by varying the QCD renormalization and factorization scales, and the PDF set following the PDF4LHC recommendations~\cite{NNPDF, PDF4LHC, Botje:2011sn, CT10}. These are found to be 4.5\% and 25\% for the $\PQq\PAQq\to \PZ \PZ$ and $\Pg\Pg\to \PZ \PZ$ backgrounds, respectively~\cite{Chatrchyan:2013mxa}. The systematic uncertainties in the reducible background estimate for the $4\Pe$, $4\mu$, and $2\Pe2\mu$ final states are determined to be 20\%, 40\%, and 25\%, respectively~\cite{Chatrchyan:2013mxa}.  In the case of the differential measurements, uncertainties in the irreducible background rates are computed for each bin, while uncertainties in the reducible background rates are assumed to be identical in all bins of the considered observable. The absolute integrated luminosity of the $\Pp \Pp$ collisions at 7 and 8\TeV has been determined with a relative precision of 2.2\%~\cite{CMS:2012rua} and 2.6\%~\cite{CMS-PAS-LUM-13-001}, respectively. For all cross section measurements, an uncertainty in the resolution of the signal mass peak of 20\% is included in the signal determination~\cite{Chatrchyan:2013mxa}.

When measuring the differential cross section as a function of the jet multiplicity, the systematic uncertainty in the jet energy scale is included as fully correlated between the signal and background estimations. This uncertainty ranges from $3\%$ for low jet multiplicity bins to $12\%$ for the highest jet multiplicity bin for the signal, and from 2\% to 16\% for background. The uncertainties related to the jet identification efficiency and the jet energy resolution are found to be negligible with respect to the jet energy scale systematic uncertainty.

The underlying assumption on the signal model used to extract the fiducial cross sections introduces an additional systematic effect on the measurement result. This effect is estimated by extracting the fiducial cross sections from data assuming a range of alternative signal models.  The alternative models include models with an arbitrary fraction of the SM Higgs boson production modes, models of Higgs-like resonances with anomalous interactions with a pair of neutral gauge bosons, or models of Higgs-like resonances with exotic decays to the four-lepton final state. These exotic models are briefly introduced in Section~\ref{sec:DatasetsSimulationPAS} and detailed in Ref.~\cite{Khachatryan:2014kca}. The largest deviation between the fiducial cross sections measured assuming these alternative signal models and the fiducial cross section measured under the SM Higgs boson assumption is quoted as the systematic effect associated with the model dependence. If we neglect the existing experimental constraints~\cite{CMS:2014ega, Khachatryan:2014kca} on the exotic signal models, the effect is found to be up to 7\% in all reported measurements, except in the case of the jet multiplicity differential measurement where in some bins the effect can be as large as 25\%. If we impose experimental constraints~\cite{CMS:2014ega, Khachatryan:2014kca} on the allowed exotic signal models, the systematic effect associated with the model dependence reduces to 3-5\% for the jet multiplicity differential measurement, and it is smaller than 1\% for the other measurements. The more conservative case which does not take into account existing experimental constraints is used to report a separate systematic uncertainty due to the model dependence.

The effect on the cross section measurement due to $m_{\PH}$ being fixed in the fit procedure is estimated from simulation to be about 1\%. The additional uncertainty due to this effect is negligible with respect to the other systematic uncertainties, and is not included  in the measurements. The overview of the main systematic effects in the case of the $\Hllll$ measurements is presented in Table~\ref{tab:SystOverview}.

\begin{table}
\centering
\topcaption{
Overview of main sources of the systematic uncertainties in the $\Hllll$ cross section measurements. More details, including the definition of the model dependence are presented in the text.
\label{tab:SystOverview}
}
\begin{tabular}{lc}
\hline
\multicolumn{2}{c}{Summary of relative systematic uncertainties} \\
\hline
\multicolumn{2}{c}{}  \\[-0.35cm]
\multicolumn{2}{c}{Common experimental uncertainties} \\
\hline
Luminosity & 2.2\% (7\TeV), 2.6\% (8\TeV) \\
Lepton identification/reconstruction efficiencies & \phantom{0}4--10\% \\
\hline
\multicolumn{2}{c}{}  \\[-0.35cm]
\multicolumn{2}{c}{Background related uncertainties} \\
\hline
QCD scale ($\PQq\PAQq\to\PZ\PZ$, $\Pg\Pg\to\PZ\PZ$) & \phantom{0}3--24\% \\
PDF set ($\PQq\PAQq\to\PZ\PZ$, $\Pg\Pg\to\PZ\PZ$) & 3--7\% \\
Reducible background ($\PZ+\mathrm{X}$) & 20--40\% \\
Jet resolution and energy scale & \phantom{0}2--16\% \\
\hline
\multicolumn{2}{c}{}  \\[-0.35cm]
\multicolumn{2}{c}{Signal related uncertainties} \\
\hline
Lepton energy scale & 0.1--0.3\% \\
Lepton energy resolution & 20\% \\
Jet energy scale and resolution & \phantom{0}3--12\% \\
\hline
\multicolumn{2}{c}{}  \\[-0.35cm]
\multicolumn{2}{c}{Combinatorial signal-induced contribution} \\
\hline
Effect on the final measurement & \phantom{0}4--11\% \\
\hline
\multicolumn{2}{c}{}  \\[-0.35cm]
\multicolumn{2}{c}{Model dependence} \\
\hline
With exp. constraints on production modes and exotic models & 1--5\% \\
No exp. constraints on production modes and exotic models & \phantom{0}7--25\% \\
\hline
\end{tabular}
\normalsize
\end{table}

\section{Results} \label{sec:Results}
\begin{figure}[htb]
\centering
      \includegraphics[width=0.48\textwidth]{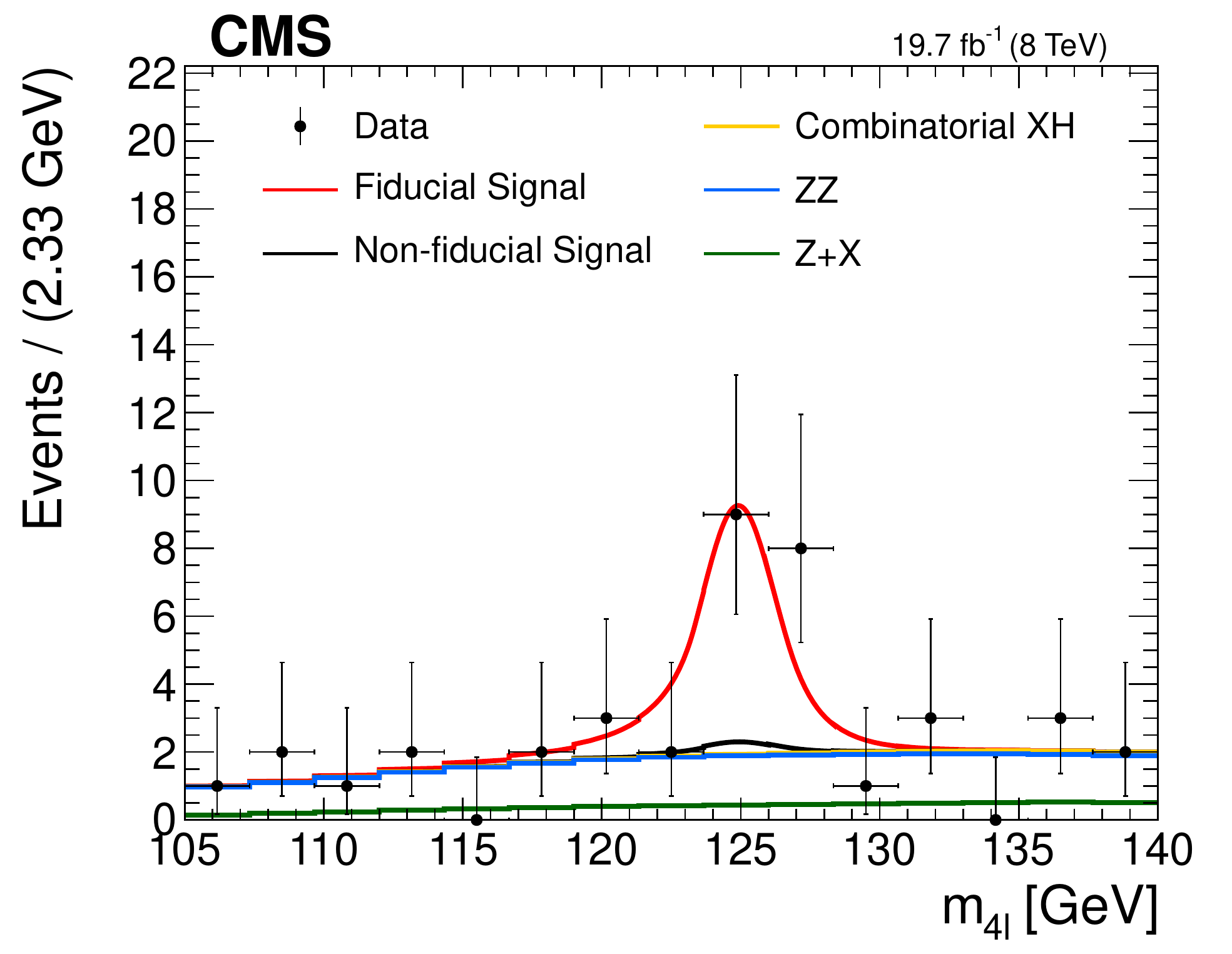}
      \label{fig:sigfit-inclusive:a}
      \includegraphics[width=0.48\textwidth]{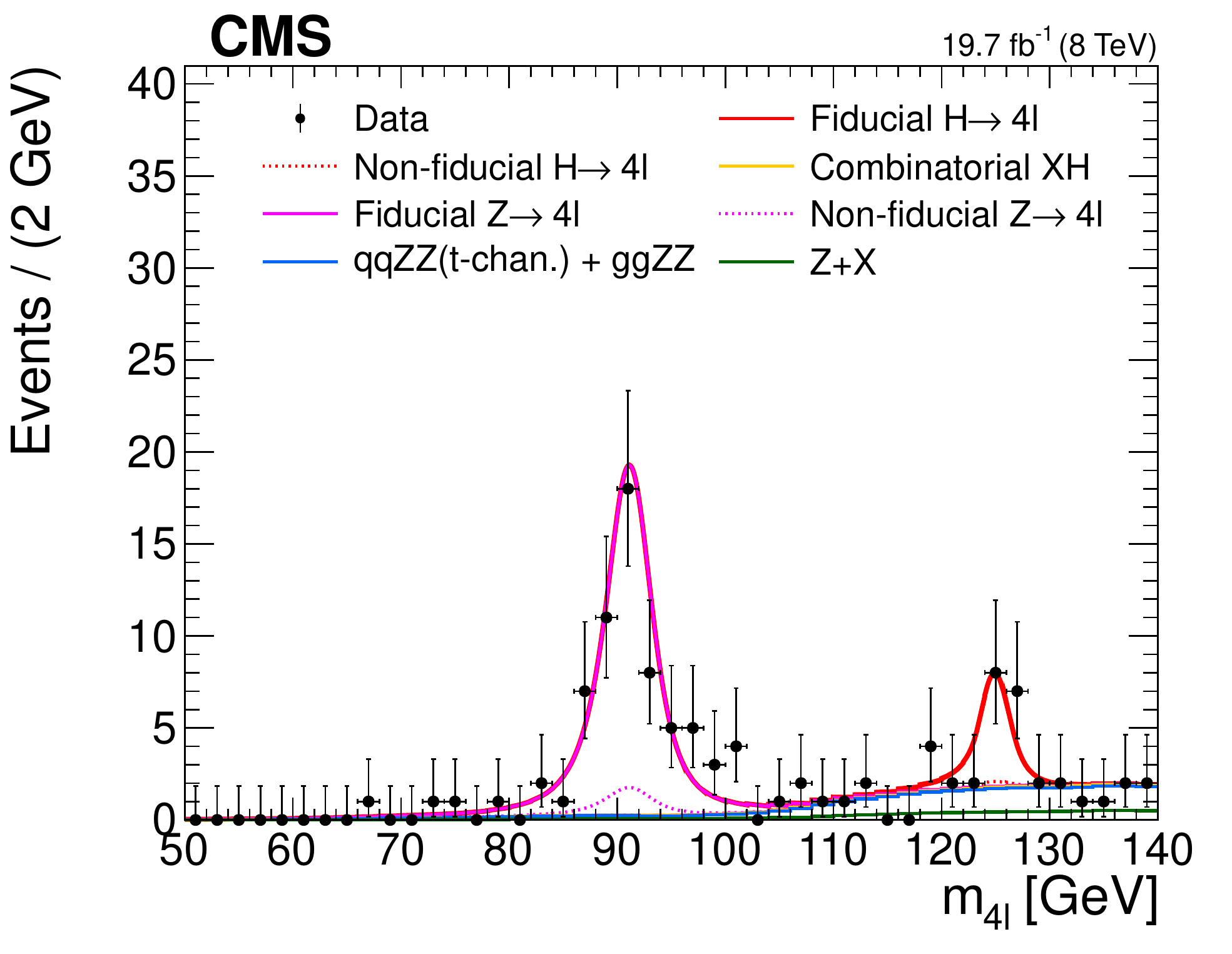}
      \label{fig:sigfit-inclusive:b}
    \caption{Observed inclusive four-lepton mass distribution and the resulting fits of the signal and background models, presented in Section~\ref{sec:Methodology}, in case of an independent $\Hllll$ fit (left) and a simultaneous $\Hllll$ and $\Zllll$ fit (right). The $\Pg\Pg \to \PH \to 4\ell$ process is modelled using \textsc{Powheg+JHUGen},
    while $\PQq \PAQq \to 4\ell$ process is modelled using \POWHEG (both $s$- and $t$/$u$-channels).
    The sub-dominant component of the Higgs boson production is denoted as XH = VBF + \VH + $\ttH$.
    }
  \label{fig:sigfit-inclusive}
\end{figure}

The result of the maximum likelihood fit to the signal and background $m_{4\ell}$ spectra in data collected at $\sqrt{s}=8\TeV$, used to extract the integrated $\Hllll$ fiducial cross section for the $m_{4\ell}$ range from 105 to 140\GeV, is shown in Fig.~\ref{fig:sigfit-inclusive} (left).  Similarly, the result of the maximum likelihood fit for the $\Hllll$ and $\Zllll$ contributions to the inclusive $m_{4\ell}$ spectra in the range from 50 to 140\GeV is shown in Fig.~\ref{fig:sigfit-inclusive} (right).

\begin{figure}[!h!tb]
\centering
    \includegraphics[width=0.65\textwidth]{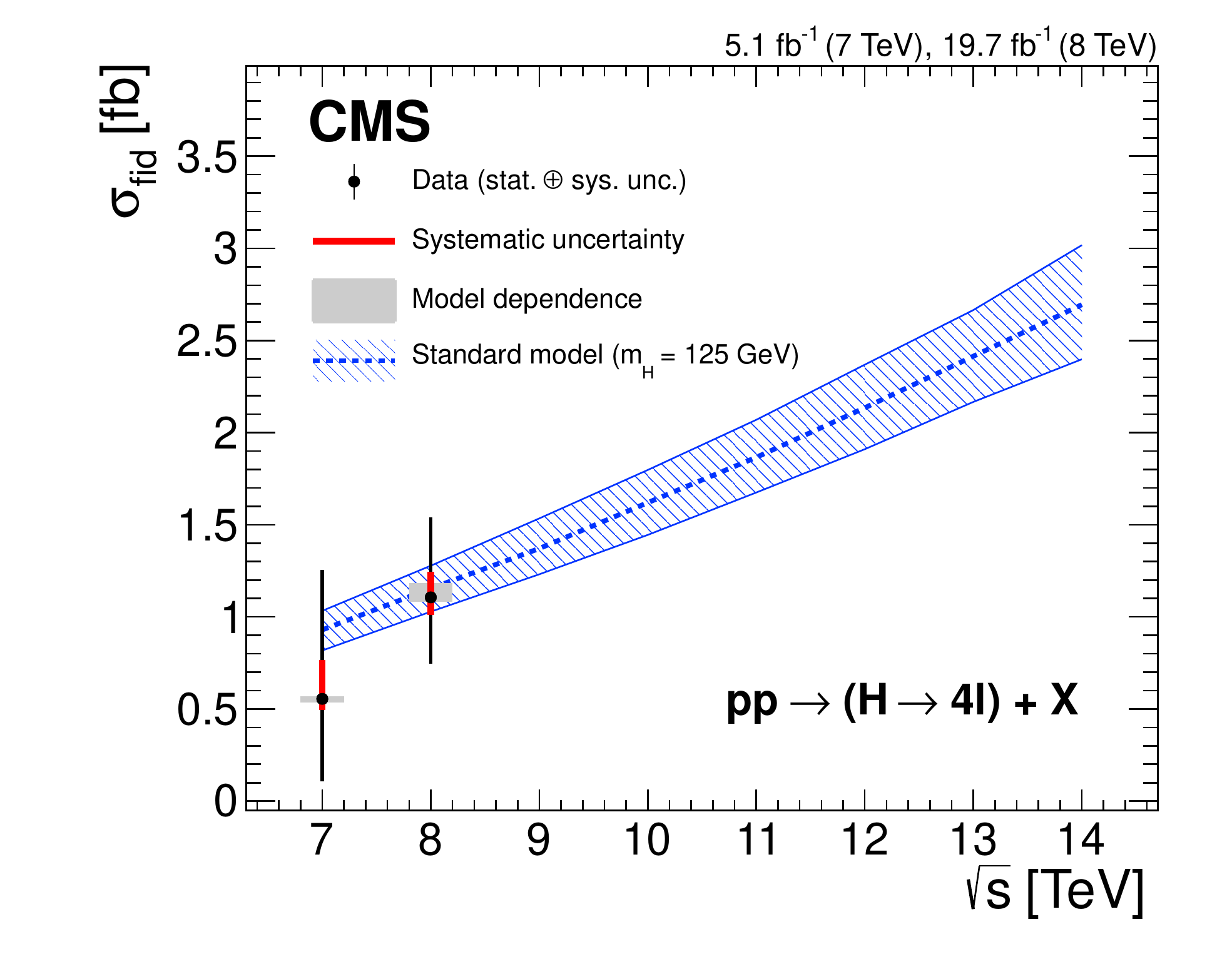}
    \caption{Results of measurements of the integrated $\Hllll$ fiducial cross section in pp collisions at 7 and 8\TeV, with a comparison to SM estimates. The red error bar represents the systematic uncertainty, while the black error bar represents the combined statistical and systematic uncertainties, summed in quadrature. The additional systematic effect associated with model dependence is represented by grey boxes.
        The theoretical estimates at NNLL+NNLO accuracy and the corresponding systematic uncertainties are shown in blue as a function of the centre-of-mass energy. The acceptance of the dominant $\ggH$ contribution is modelled at the parton level using \textsc{HRes}, and corrected for hadronization and underlying-event effects estimated using \textsc{Powheg+JHUGen} and \textsc{Pythia 6.4}.
}
  \label{fig:inclusive-results}
\end{figure}

Individual measurements of integrated $\Hllll$ fiducial cross sections at 7 and 8\TeV, performed in the $m_{4\ell}$ range from 105 to 140\GeV, are presented in Table~\ref{tab:incresultsPAS} and Fig.~\ref{fig:inclusive-results}.
The central values of the measurements are obtained assuming the SM Higgs boson signal with $m_{\PH} =125.0\GeV$, modelled by the \textsc{ Powheg+JHUGen} for the $\ggH$ contribution, \POWHEG for the VBF contribution, and \PYTHIA for the \VH + $\ttH$ contributions. In Table~\ref{tab:incresultsPAS} and hereafter, the sub-dominant component of the signal is denoted as XH = VBF + \VH + $\ttH$.

\begin{table}[!h!tb]
\renewcommand{\arraystretch}{1.2}
\centering
      \topcaption{
Results of the $\Hllll$ integrated fiducial cross section measurements performed in the $m_{4\ell}$ range from 105 to 140\GeV for pp collisions at 7 and 8\TeV, and comparison to the theoretical estimates obtained at NNLL+NNLO accuracy. Statistical and systematic uncertainties, as well as the model-dependent effects are quoted separately.
The sub-dominant component of the Higgs boson production is denoted as XH = VBF + \VH + $\ttH$.
        } \label{tab:incresultsPAS}
\begin{tabular}{c|c}
\hline
\multicolumn{2}{c}{}  \\[-0.4cm]
\multicolumn{2}{c}{Fiducial cross section $\Hllll$ at 7\TeV} \\
\hline
Measured & $0.56^{+0.67}_{-0.44}\stat\;{}^{+0.21}_{-0.06}\syst\;{}\pm{0.02}\,\text{(model)} \unit{fb}$  \\
$\ggH$(\textsc{HRes}) + XH & $0.93^{+0.10}_{-0.11}\unit{fb}$ \\
\hline
\multicolumn{2}{c}{}  \\[-0.4cm]
\multicolumn{2}{c}{Fiducial cross section $\Hllll$ at 8\TeV} \\
\hline
Measured & $1.11^{+0.41}_{-0.35}\stat\;{}^{+0.14}_{-0.10}\syst\;{}^{+0.08}_{-0.02}\,\text{(model)}\unit{fb}$  \\
$\ggH$(\textsc{HRes}) + XH & $1.15^{+0.12}_{-0.13}\unit{fb}$ \\
\hline
\multicolumn{2}{c}{}  \\[-0.4cm]
\multicolumn{2}{c}{Ratio of $\Hllll$ fiducial cross sections at 7 and 8\TeV} \\
\hline
Measured & $0.51^{+0.71}_{-0.40}\stat\;{}^{+0.13}_{-0.05}\syst\;{}^{+0.00}_{-0.03}\,\text{(model)}$  \\
$\ggH$(\textsc{Hres}) + XH & $0.805^{+0.003}_{-0.010}$ \\
\hline
\end{tabular}
\end{table}

The measured fiducial cross sections are compared to the SM NNLL+NNLO theoretical estimations in which the acceptance of the dominant $\ggH$ contribution is modelled using \textsc{Powheg+JHUGen}, \textsc{MiNLO HJ}, or \textsc{HRes},
as discussed in Section~\ref{sec:DatasetsSimulationPAS}.
The total uncertainty in the NNLL+NNLO theoretical estimates is computed according to Ref.~\cite{Heinemeyer:2013tqa}, and includes uncertainties due to the QCD renormalization and factorization scales ($\sim$7.8\%), PDFs and strong coupling constant $\alpha_\mathrm{S}$ modelling ($\sim$7.5\%), as well as the acceptance ($2\%$) and branching fraction ($2\%$) uncertainties. In the computation of the total uncertainty the PDFs/$\alpha_\mathrm{S}$ uncertainties are assumed to be correlated between the VBF and \VH production modes (dominantly quark-antiquark initiated), and anticorrelated between the $\ggH$ and $\ttH$ production modes (dominantly gluon-gluon initiated). Furthermore, the QCD scale uncertainties are considered to be uncorrelated, while uncertainties in the acceptance and branching fraction are considered to be correlated across all production modes.
The differences in how the \textsc{Powheg+JHUGen}, \textsc{MiNLO HJ}, and \textsc{HRes} generators model the acceptance of the $\ggH$ contribution are found to be an order of magnitude lower than the theoretical uncertainties, and in Table~\ref{tab:incresultsPAS} and Fig.~\ref{fig:inclusive-results} we show estimations obtained using \textsc{HRes}.

The measured $\Hllll$ fiducial cross section at 8\TeV is found to be in a good agreement with the theoretical estimations within the associated uncertainties. The uncertainty of the measurement is largely dominated by its statistical component of about 37\%, while the systematic component is about 12\%. The theoretical uncertainty of about 11\% is comparable to the systematic uncertainty, and is larger than the model dependence of the extracted results, which is about 7\%. In the case of the cross section at 7\TeV, as well as the ratio of cross sections at 7 and 8\TeV, the measured cross sections are lower but still in agreement with the SM theoretical estimations within the large statistical uncertainties.

\begin{table}[!h!tb]
\renewcommand{\arraystretch}{1.2}
\centering
      \topcaption{
The $\Zllll$ integrated fiducial cross section at 8\TeV in the $m_{4\ell}$ range from 50 to 105\GeV,
and the ratio of 8\TeV fiducial cross sections of $\Hllll$ and $\Zllll$ obtained
from a simultaneous fit of mass peaks of $\Zllll$ and $\Hllll$ in the mass window 50 to 140\GeV.
The sub-dominant component of the Higgs boson production is denoted as XH = VBF + \VH + $\ttH$.
        } \label{tab:incresultsZ4l}
\begin{tabular}{c|c}
\hline
\multicolumn{2}{c}{}  \\[-0.4cm]
\multicolumn{2}{c}{Fiducial cross section $\Zllll$ at 8\TeV} \\[-0.15cm]
\multicolumn{2}{c}{($50 < m_{4\ell} < 105\GeV$)} \\
\hline
Measured & $4.81^{+0.69}_{-0.63} \stat\;{}^{+0.18}_{-0.19}\syst \unit{fb}$  \\
\POWHEG & $ 4.56\pm{0.19}\unit{fb}$ \\
\hline
\multicolumn{2}{c}{}  \\[-0.4cm]
\multicolumn{2}{c}{Ratio of fiducial cross sections of $\Hllll$ and $\Zllll$ at 8\TeV} \\[-0.15cm]
\multicolumn{2}{c}{($50 < m_{4\ell} < 140\GeV$)} \\
\hline
Measured & $0.21^{+0.09}_{-0.07} \stat \pm{0.01} \syst $  \\
$\ggH$(\textsc{HRes}) + XH and $\Zllll$ (\POWHEG) & $ 0.25\pm{0.04}$ \\
\hline
\end{tabular}
\end{table}

The result of the measurement of the integrated $\Zllll$ fiducial cross section at 8\TeV in the $m_{4\ell}$ range from 50 to 105\GeV is summarized in Table~\ref{tab:incresultsZ4l}.
The measured $\Zllll$ cross section is found to be in good agreement with the theoretical estimations obtained using \POWHEG.
As the total relative uncertainty in the $\Zllll$ measurement is about 2.6 times lower than the relative uncertainty in the $\Hllll$ measurement,
the good agreement between the measured and estimated $\Zllll$ cross section provides a validation of the measurement procedure in data.

In addition, a simultaneous fit for the $\Hllll$ and $\Zllll$ resonances is performed in the $m_{4\ell}$ range from 50 to 140\GeV, and the ratio of the corresponding fiducial cross sections is extracted.
The measurement of the ratio of these cross sections, when masses of the two resonances are fixed in the fit, is presented in Table~\ref{tab:incresultsZ4l}. A good agreement between the measured ratio and its SM theoretical
estimation is observed.
In the scenario in which the masses of the two resonances are allowed to vary, as discussed in Section~\ref{sec:Methodology}, the fitted value for the mass difference between the two resonances is found to be $\Delta m = m_{\PH} - m_{\PZ} = 34.2\pm{0.7}\GeV$.
As discussed in Ref.~\cite {Roinishvili:2013goa}, it is worth noting that by using the measured mass difference $\Delta m$ and the PDG value of the $\PZ$ boson mass $m_{\PZ}^{\rm{PDG}}$ which is precisely determined in other experiments, the Higgs boson mass can be extracted as $m_{\rm{H}} = m_{\PZ}^{\rm{PDG}} + \Delta m = 125.4 \pm 0.7\GeV$.
This result is in agreement with the best fit value for $m_{\PH}$ obtained from the dedicated mass measurement in this final state~\cite{Chatrchyan:2013mxa}, and provides further validation of the measurement procedure.

\begin{figure}[h]
\centering
      \includegraphics[width=0.46\textwidth]{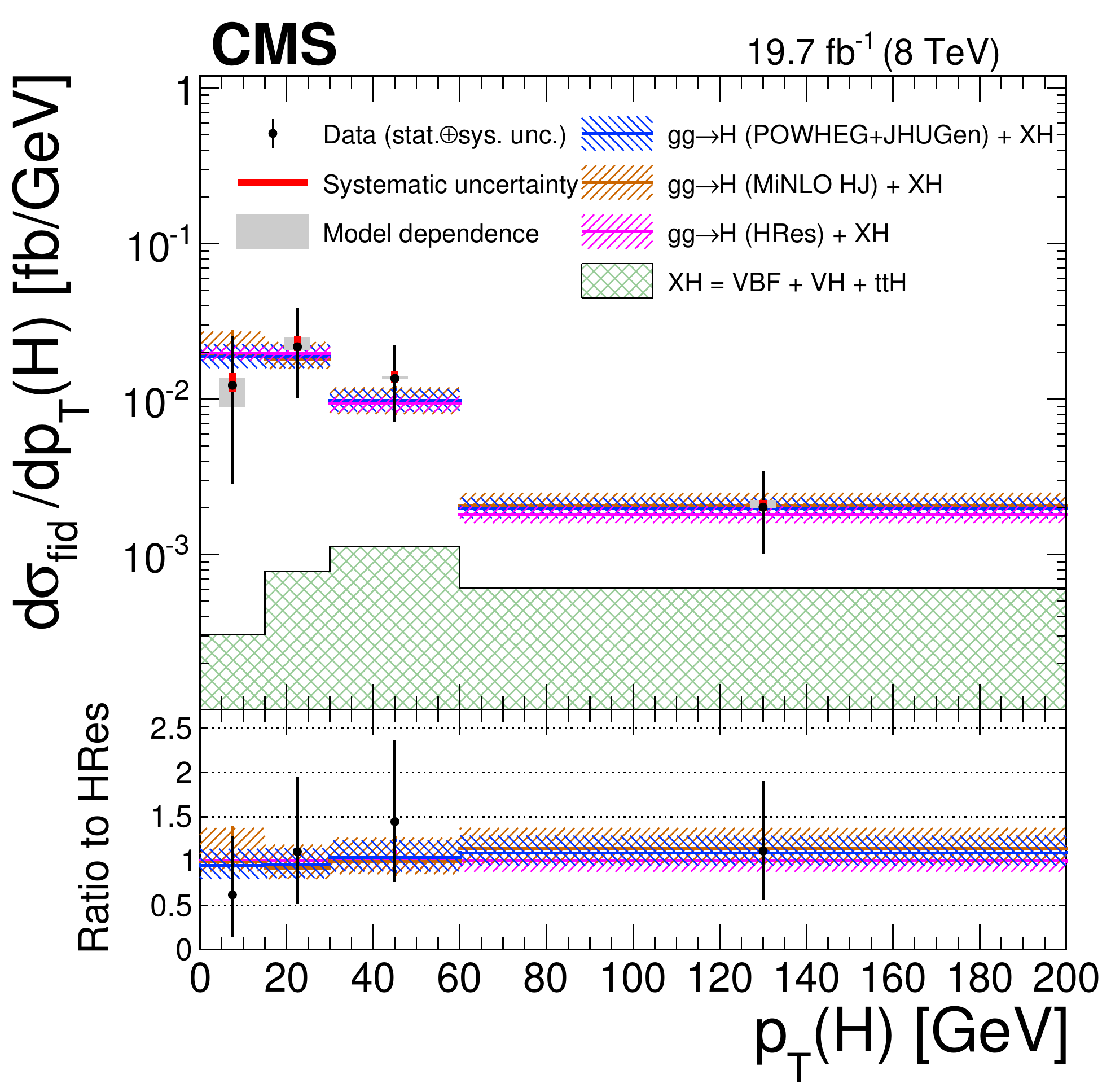}
      \includegraphics[width=0.46\textwidth]{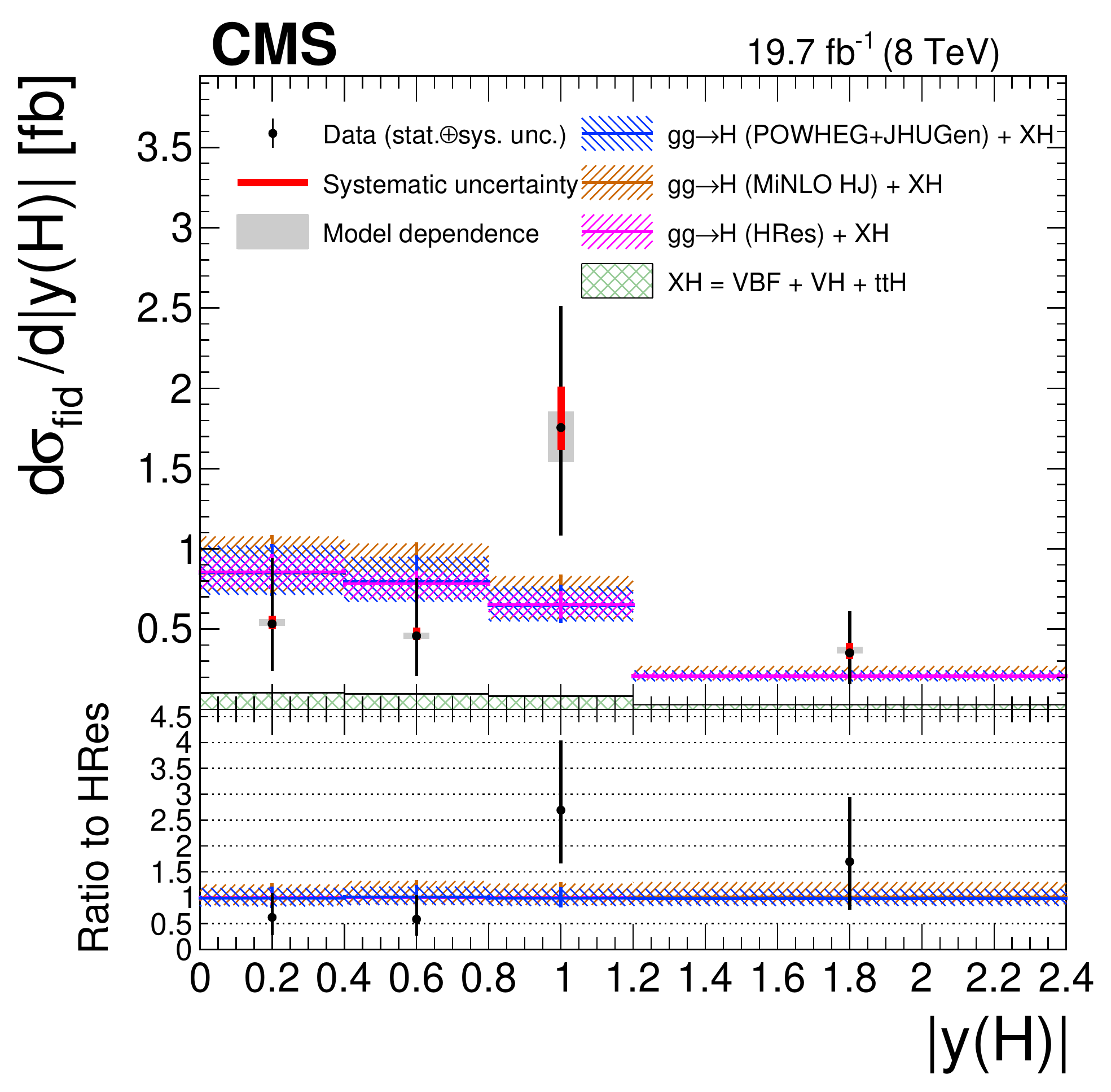}
    \caption{Results of the differential $\Hllll$ fiducial cross section measurements and comparison to the theoretical estimates for the transverse momentum (left) and the rapidity (right) of the four-lepton system. The red error bars represent the systematic uncertainties, while black error bars represent the combined statistical and systematic uncertainties, summed in quadrature. The additional systematic uncertainty associated with the model dependence is separately represented by the grey boxes. Theoretical estimates, in which the acceptance of the dominant $\ggH$ contribution is modelled by \textsc{Powheg+JHUGen}+\PYTHIA, \textsc{Powheg MiNLO HJ}+\PYTHIA, and \textsc{HRes} generators as discussed in Section~\ref{sec:DatasetsSimulationPAS}, are shown in blue, brown, and pink, respectively. The sub-dominant component of the signal XH is indicated separately in green. In all estimations the total cross section is normalized to the SM estimate computed at NNLL+NNLO accuracy. Systematic uncertainties correspond to the accuracy of the generators used to derive the differential estimations. The bottom panel shows the ratio of data or theoretical estimates to the \textsc{HRes} theoretical estimations.
    }
  \label{fig:differential-results}
\end{figure}

\begin{figure}[!h]
\centering
      \includegraphics[width=0.43\textwidth]{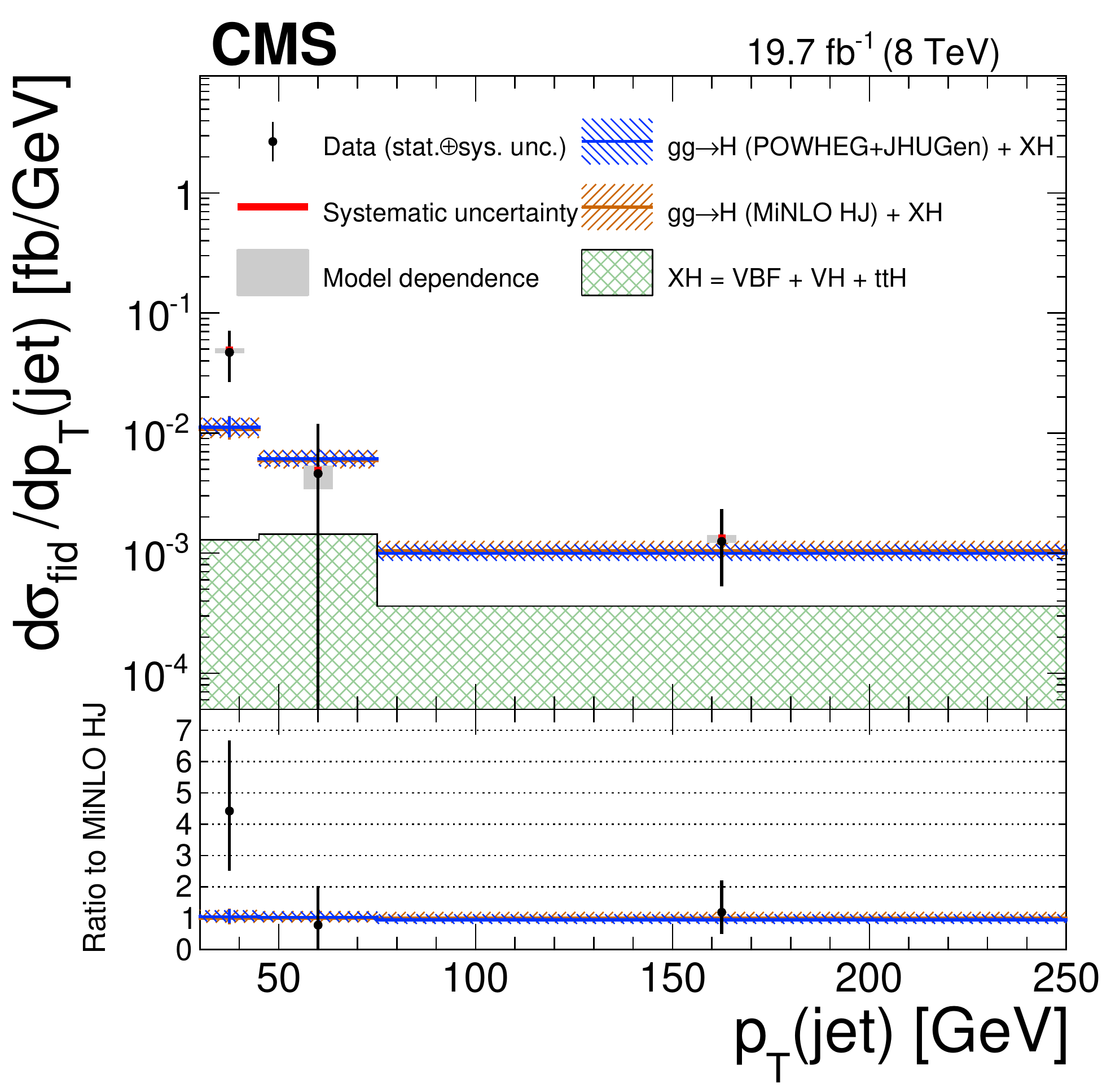}
      \includegraphics[width=0.43\textwidth]{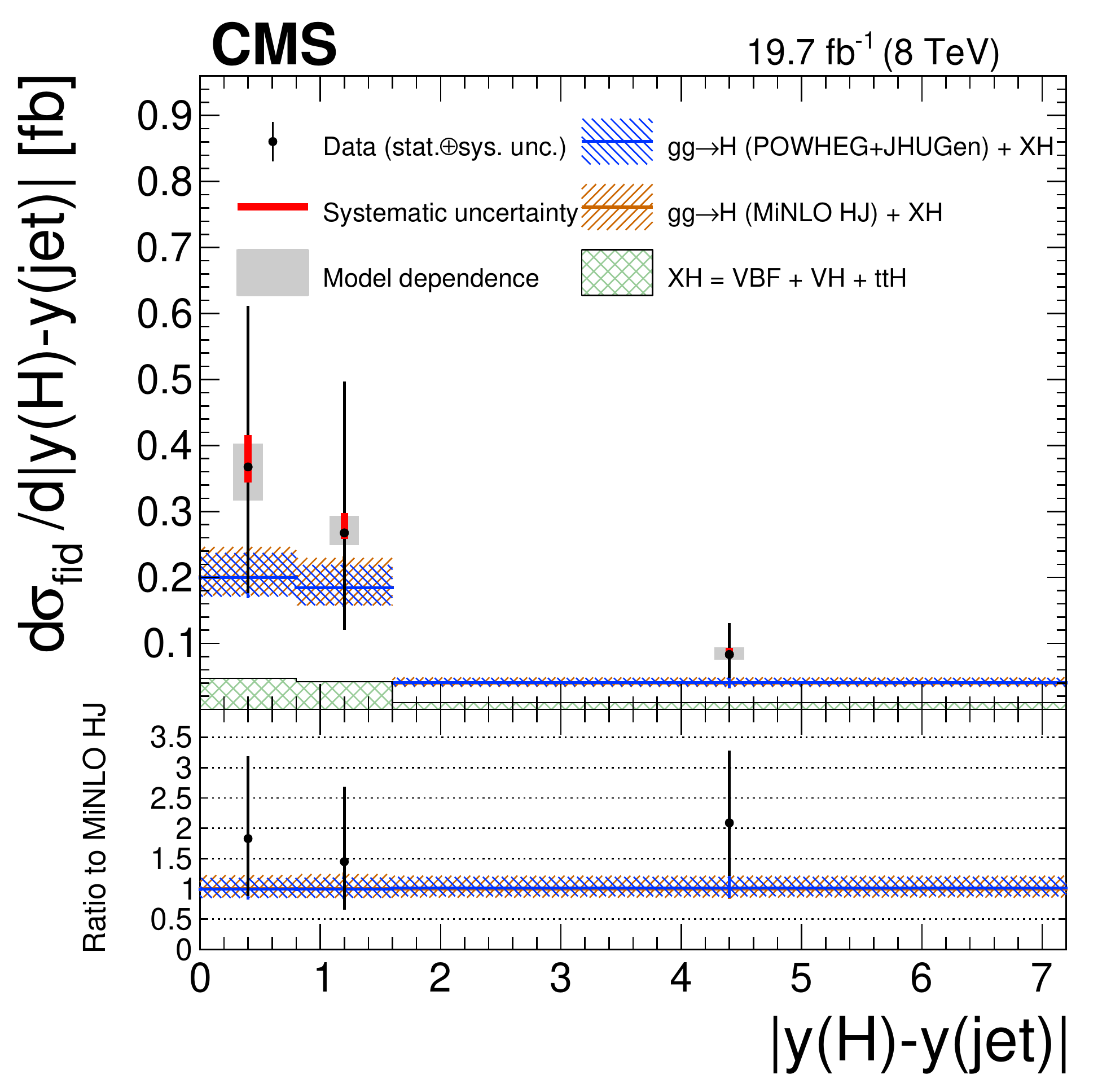}
\\
      \includegraphics[width=0.43\textwidth]{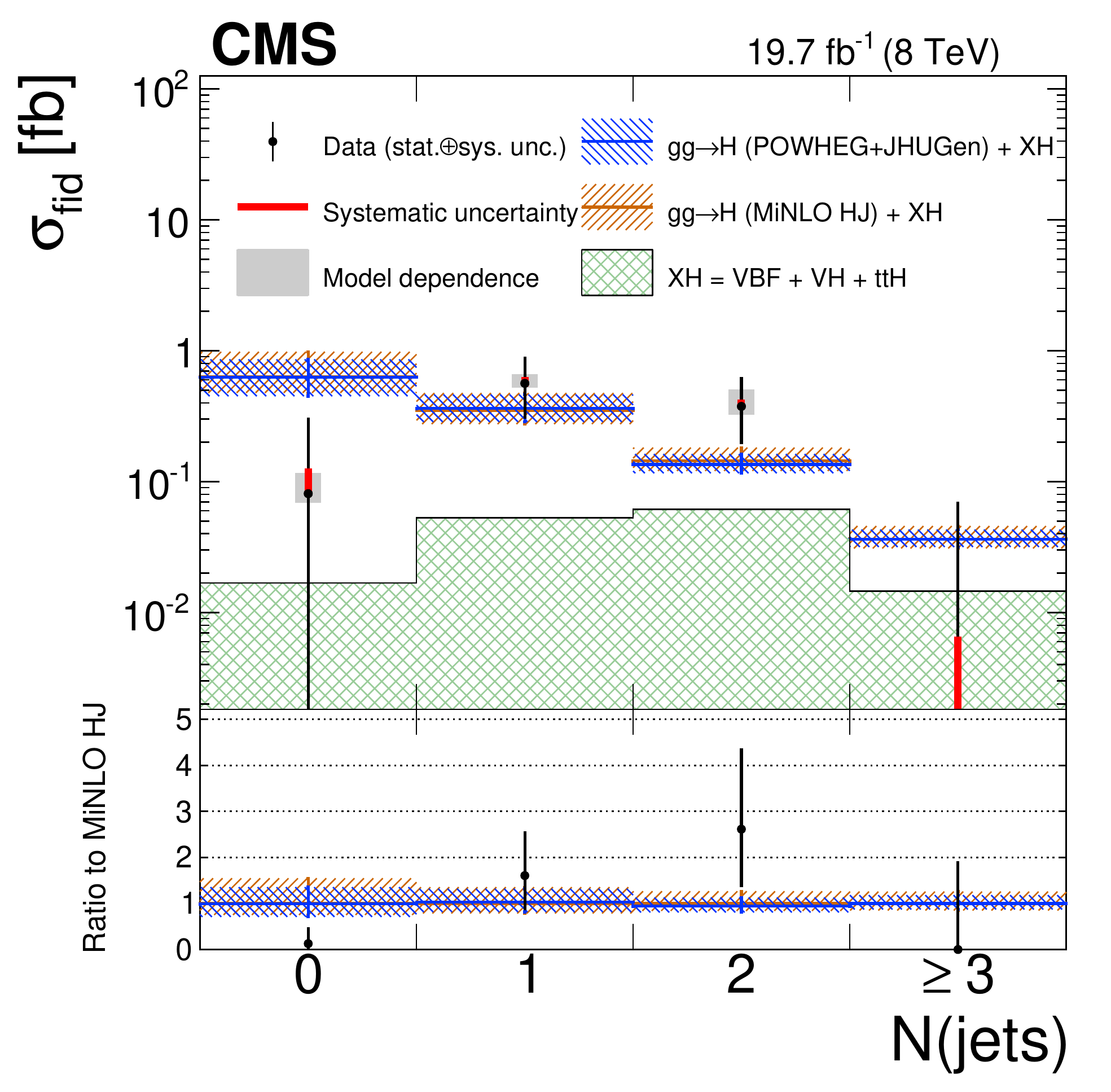}
    \caption{Results of the differential $\Hllll$ fiducial cross section measurements and comparison to the theoretical estimates for the transverse momentum of the leading jet (top left), separation in rapidity between the Higgs boson candidate and the leading jet (top right), as well as for the jet multiplicity (bottom). The red error bars represent the systematic uncertainties, while black error bars represent the combined statistical and systematic uncertainties, summed in quadrature. The additional systematic uncertainty associated with the model dependence is separately represented by the grey boxes. Theoretical estimations, in which the acceptance of the dominant $\ggH$ contribution is modelled by \textsc{Powheg+JHUGen+}\PYTHIA, and \textsc{Powheg MiNLO HJ+}\PYTHIA generators, as discussed in Section~\ref{sec:DatasetsSimulationPAS}, are shown in blue and brown, respectively. The sub-dominant component of the signal XH is indicated separately in green. In all estimations the total cross section is normalized to the SM estimate computed at NNLL+NNLO accuracy. Systematic uncertainties correspond to the accuracy of the generators used to derive the differential estimations. The bottom panel shows the ratio of data or theoretical estimates to the \textsc{Powheg MiNLO HJ} theoretical estimations.
    }
  \label{fig:differential-results-2}
\end{figure}

The measured differential $\Hllll$ cross sections at 8\TeV, along with the theoretical estimations for a SM Higgs boson with $m_{\PH}=125.0\GeV$ are presented in Figs.~\ref{fig:differential-results} and \ref{fig:differential-results-2}. Results of the measurements are shown for the transverse momentum and the rapidity of the four-lepton system, jet multiplicity, transverse momentum of the leading jet, as well as separation in rapidity between the Higgs boson candidate and the leading jet. The uncertainty in the theoretical estimation for the dominant $\ggH$ process is computed in each bin of the considered observable by the generator used for the particular signal description (\textsc{Powheg+JHUGen}, \textsc{Powheg MiNLO HJ}, or \textsc{HRes}). The theoretical uncertainties for the associated production mechanisms are taken as constant across the bins of the differential observables and are obtained from Ref.~\cite{Heinemeyer:2013tqa}.

The measurement of the transverse momentum of the four-lepton system probes the perturbative QCD calculations of the dominant loop-mediated $\ggH$ production mechanism, in which the transverse momentum $\pt(\PH)$ is expected to be balanced by the emission of soft gluons and quarks. In addition, the rapidity distribution of the four-lepton system, $y(\PH)$ is sensitive both to the modelling of the gluon fusion production mechanism and to the PDFs of the colliding protons. The measured differential cross sections for these two observables are shown in Fig.~\ref{fig:differential-results}. Results are compared to the theoretical estimations in which the dominant $\ggH$ contribution is modelled using \textsc{Powheg+JHUGen}, \textsc{Powheg MiNLO HJ}, and \textsc{HRes}.  In case of the \textsc{HRes}, the $\ggH$ acceptance is modelled at the parton level, and corrected for the hadronization and underlying event effects in bins of the considered differential observable, as discussed in Section~\ref{sec:DatasetsSimulationPAS}. The observed distributions are compatible with the SM-based theoretical estimations within the large associated uncertainties.

Similarly, the jet multiplicity $N(\text{jets})$, transverse momentum of the leading jet $\pt(\text{jet})$, and its separation in rapidity from the Higgs boson candidate $\abs{y(\PH) - y(\text{jet})}$ are sensitive to the theoretical modelling of hard quark and gluon radiation in this process, as well as to the relative contributions of different Higgs boson production mechanisms. The measured differential cross sections for the leading jet transverse momentum, and its separation in rapidity from the Higgs boson candidate are shown in Fig.~\ref{fig:differential-results-2}, and are found to be compatible with the SM-based estimations within the large uncertainties. In the case of the jet multiplicity cross section, also shown in Fig.~\ref{fig:differential-results-2}, we observe the largest deviation from the SM-based estimations. The $p$-value that quantifies the compatibility of the jet multiplicity distribution between data and SM estimations is $p=0.13$. It is computed from the difference between the $-2\log(\mathcal{L})$ at its best fit value and the value with the cross sections fixed to the theoretical estimation based on the \textsc{ Powheg+JHUGen} description of the $\ggH$ process. Furthermore, we have performed the measurement of the differential $\Zllll$ cross sections at 8\TeV for the same set of observables used in the $\Hllll$ measurements, including the jet multiplicity, and have found a good agreement with the theoretical estimations. The $p$-values for the differential distributions of $\Zllll$ events range from 0.21 in case of rapidity of the Z boson, to 0.99 for some of the angles defined by the four leptons in the Collins-Soper reference frame~\cite{Collins:1977iv}. As the relative statistical uncertainty in the $\Zllll$ measurement is lower than the relative uncertainty in the $\Hllll$ measurement, these results provide additional validation of the measurement procedure in data.

\section{Summary}  \label{sec:ConclusionPAS}
We have presented measurements of the integrated and differential fiducial cross sections for the production of four leptons via the $\PH \to 4\ell$ decays in pp collisions at centre-of-mass energies of 7 and 8\TeV. The measurements were performed using collision data corresponding to integrated luminosities of \usedLumiA at 7\TeV and \usedLumiB at 8\TeV. The differential cross sections were measured as a function of the transverse momentum and the rapidity of the four-lepton system, the transverse momentum of the leading jet, the difference in rapidity between the Higgs boson candidate and the leading jet, and the jet multiplicity. Measurements of the fiducial cross section for the production of four leptons via the $\Zllll$ decays, as well as its ratio to the $\Hllll$ cross section, were also performed using the 8\TeV data. The uncertainty in the measurements due to the assumptions in the model of Higgs boson properties was estimated by studying a range of exotic Higgs boson production and spin-parity models. It was found to be lower than 7\% of the fiducial cross section. The integrated fiducial cross section for the four leptons production via the $\PH \to 4\ell$ decays is measured to be $0.56^{+0.67}_{-0.44}\stat\;{}^{+0.21}_{-0.06}\syst\unit{fb}$ and $1.11^{+0.41}_{-0.35}\stat\;{}^{+0.14}_{-0.10}\syst\unit{fb}$ at 7 and 8\TeV, respectively. The measurements are found to be compatible with theoretical calculations based on the standard model.

\section*{Acknowledgements}
\hyphenation{Bundes-ministerium Forschungs-gemeinschaft Forschungs-zentren} We congratulate our colleagues in the CERN accelerator departments for the excellent performance of the LHC and thank the technical and administrative staffs at CERN and at other CMS institutes for their contributions to the success of the CMS effort. In addition, we gratefully acknowledge the computing centres and personnel of the Worldwide LHC Computing Grid for delivering so effectively the computing infrastructure essential to our analyses. Finally, we acknowledge the enduring support for the construction and operation of the LHC and the CMS detector provided by the following funding agencies: the Austrian Federal Ministry of Science, Research and Economy and the Austrian Science Fund; the Belgian Fonds de la Recherche Scientifique, and Fonds voor Wetenschappelijk Onderzoek; the Brazilian Funding Agencies (CNPq, CAPES, FAPERJ, and FAPESP); the Bulgarian Ministry of Education and Science; CERN; the Chinese Academy of Sciences, Ministry of Science and Technology, and National Natural Science Foundation of China; the Colombian Funding Agency (COLCIENCIAS); the Croatian Ministry of Science, Education and Sport, and the Croatian Science Foundation; the Research Promotion Foundation, Cyprus; the Ministry of Education and Research, Estonian Research Council via IUT23-4 and IUT23-6 and European Regional Development Fund, Estonia; the Academy of Finland, Finnish Ministry of Education and Culture, and Helsinki Institute of Physics; the Institut National de Physique Nucl\'eaire et de Physique des Particules~/~CNRS, and Commissariat \`a l'\'Energie Atomique et aux \'Energies Alternatives~/~CEA, France; the Bundesministerium f\"ur Bildung und Forschung, Deutsche Forschungsgemeinschaft, and Helmholtz-Gemeinschaft Deutscher Forschungszentren, Germany; the General Secretariat for Research and Technology, Greece; the National Scientific Research Foundation, and National Innovation Office, Hungary; the Department of Atomic Energy and the Department of Science and Technology, India; the Institute for Studies in Theoretical Physics and Mathematics, Iran; the Science Foundation, Ireland; the Istituto Nazionale di Fisica Nucleare, Italy; the Ministry of Science, ICT and Future Planning, and National Research Foundation (NRF), Republic of Korea; the Lithuanian Academy of Sciences; the Ministry of Education, and University of Malaya (Malaysia); the Mexican Funding Agencies (CINVESTAV, CONACYT, SEP, and UASLP-FAI); the Ministry of Business, Innovation and Employment, New Zealand; the Pakistan Atomic Energy Commission; the Ministry of Science and Higher Education and the National Science Centre, Poland; the Funda\c{c}\~ao para a Ci\^encia e a Tecnologia, Portugal; JINR, Dubna; the Ministry of Education and Science of the Russian Federation, the Federal Agency of Atomic Energy of the Russian Federation, Russian Academy of Sciences, and the Russian Foundation for Basic Research; the Ministry of Education, Science and Technological Development of Serbia; the Secretar\'{\i}a de Estado de Investigaci\'on, Desarrollo e Innovaci\'on and Programa Consolider-Ingenio 2010, Spain; the Swiss Funding Agencies (ETH Board, ETH Zurich, PSI, SNF, UniZH, Canton Zurich, and SER); the Ministry of Science and Technology, Taipei; the Thailand Center of Excellence in Physics, the Institute for the Promotion of Teaching Science and Technology of Thailand, Special Task Force for Activating Research and the National Science and Technology Development Agency of Thailand; the Scientific and Technical Research Council of Turkey, and Turkish Atomic Energy Authority; the National Academy of Sciences of Ukraine, and State Fund for Fundamental Researches, Ukraine; the Science and Technology Facilities Council, UK; the US Department of Energy, and the US National Science Foundation.

Individuals have received support from the Marie-Curie programme and the European Research Council and EPLANET (European Union); the Leventis Foundation; the A. P. Sloan Foundation; the Alexander von Humboldt Foundation; the Belgian Federal Science Policy Office; the Fonds pour la Formation \`a la Recherche dans l'Industrie et dans l'Agriculture (FRIA-Belgium); the Agentschap voor Innovatie door Wetenschap en Technologie (IWT-Belgium); the Ministry of Education, Youth and Sports (MEYS) of the Czech Republic; the Council of Science and Industrial Research, India; the HOMING PLUS programme of the Foundation for Polish Science, cofinanced from European Union, Regional Development Fund; the OPUS programme of the National Science Center (Poland); the Compagnia di San Paolo (Torino); the Consorzio per la Fisica (Trieste); MIUR project 20108T4XTM (Italy); the Thalis and Aristeia programmes cofinanced by EU-ESF and the Greek NSRF; the National Priorities Research Program by Qatar National Research Fund; the Rachadapisek Sompot Fund for Postdoctoral Fellowship, Chulalongkorn University (Thailand); and the Welch Foundation, contract C-1845.

\bibliography{auto_generated}

\cleardoublepage \appendix\section{The CMS Collaboration \label{app:collab}}\begin{sloppypar}\hyphenpenalty=5000\widowpenalty=500\clubpenalty=5000\textbf{Yerevan Physics Institute,  Yerevan,  Armenia}\\*[0pt]
V.~Khachatryan, A.M.~Sirunyan, A.~Tumasyan
\vskip\cmsinstskip
\textbf{Institut f\"{u}r Hochenergiephysik der OeAW,  Wien,  Austria}\\*[0pt]
W.~Adam, E.~Asilar, T.~Bergauer, J.~Brandstetter, E.~Brondolin, M.~Dragicevic, J.~Er\"{o}, M.~Flechl, M.~Friedl, R.~Fr\"{u}hwirth\cmsAuthorMark{1}, V.M.~Ghete, C.~Hartl, N.~H\"{o}rmann, J.~Hrubec, M.~Jeitler\cmsAuthorMark{1}, V.~Kn\"{u}nz, A.~K\"{o}nig, M.~Krammer\cmsAuthorMark{1}, I.~Kr\"{a}tschmer, D.~Liko, T.~Matsushita, I.~Mikulec, D.~Rabady\cmsAuthorMark{2}, B.~Rahbaran, H.~Rohringer, J.~Schieck\cmsAuthorMark{1}, R.~Sch\"{o}fbeck, J.~Strauss, W.~Treberer-Treberspurg, W.~Waltenberger, C.-E.~Wulz\cmsAuthorMark{1}
% ...
\vskip\cmsinstskip
% ...
1:~~Also at Vienna University of Technology, Vienna, Austria\\
2:~~Also at CERN, European Organization for Nuclear Research, Geneva, Switzerland\\
% ...

\end{sloppypar}
\end{document}